\documentclass[
  aps,prd,reprint,
  superscriptaddress,
  floatfix
]{revtex4-2}

\usepackage{microtype}
\usepackage{makecell} 
\usepackage{graphicx}
\usepackage{dcolumn}
\usepackage{bm}
\usepackage{caption} 
\usepackage{mathtools} 
\usepackage{tensor} 
\usepackage[hidelinks]{hyperref} 
\usepackage{amssymb}
\usepackage{mathtools}
\usepackage{mathrsfs}
\usepackage{amsmath} 
\usepackage{xcolor}
\usepackage{array} 
\usepackage{booktabs}
\usepackage{subcaption}
\usepackage{enumitem}
\usepackage{soul}
\usepackage[toc,page]{appendix}

\begin{document}

\title{On the Possibility of the Existence of Wormholes in Nature}

\author{Leonel Bixano}
    \email{Contact author: leonel.delacruz@cinvestav.mx}
\author{Tonatiuh Matos}%
 \email{Contact author: tonatiuh.matos@cinvestav.mx}
\affiliation{Departamento de F\'{\i}sica, Centro de Investigaci\'on y de Estudios Avanzados del Instituto Politécnico Nacional, Av. Instituto Politécnico Nacional 2508, San Pedro Zacatenco, M\'exico 07360, CDMX.
}%

\date{\today}

\begin{abstract}
To date, many exotic predictions of Einstein's equations have been corroborated, with one exception: wormholes (WHs). In this work, we analyse a new exact solution combination to the Einstein-Maxwell-Dilaton or Phantom equations, which represent rotating WHs with magnetic and electric fields. The solution contains a ring singularity that, like other solutions, satisfies Wormhole Cosmic Censorship; that is, the singularity is causally disconnected by the WH throat. We show that this WH is traversable through regions close to the poles, where the tidal forces and magnetic field are reasonably small. We satisfy the energy conditions for the dilaton-like solution. We argue that if dilaton-like interactions can exist in nature, for example, if extra dimensions exist as in superstring theory, WHs are a natural realistic prediction of Einstein's equations, and we should be able to observe them somehow, probably as black hole mimics.We present several examples of realistic WH that may occur in nature, and we also introduce a new exact black hole solution.
\end{abstract}

\maketitle

\section{Introduction}
Since Einstein's equations (EEq) were formulated in 1916 \cite{Einstein:1916vd}, their predictions have been stunning. Einstein himself predicted that his equations implied the existence of gravitational fluctuations, gravitational waves \cite{Einstein:1916cc}, but these are so small that, according to him, they are impossible to see or detect. Today, we have detected them hundreds of times \cite{LIGOScientific:2016aoc}, giving rise to a new era of astronomy where we detect astrophysical objects using gravitational waves instead of photons. In the same year, Karl Schwarzschild found the first exact, spherically symmetric solution to EEq \cite{Schwarzschild:1916uq}. Later, it was discovered that Schwarzschild's solution implies the existence of small, compact objects where light cannot escape \cite{Oppenheimer:1939ue}. Therefore, these objects were called black holes. At that time it was believed that the existence of these objects was impossible and that there must be a mechanism to prevent their existence. Today, we have at least two photographs of these objects \cite{EventHorizonTelescope:2019dse} and hundreds of detections of collisions of these objects using gravitational waves \cite{LIGOScientific:2016aoc}. One of the most interesting predictions of the EEq is the expansion of the universe. Even Einstein refused to accept it. He famously added the cosmological constant to his equations to prevent expansion \cite{Einstein:1917ce}, but without success. Today, we not only know that the universe is expanding \cite{Hubble:1929ig}, but that it is accelerating \cite{SupernovaSearchTeam:1998fmf}, something Einstein's equations cannot predict. And so on. To date, we are unaware of any important physical predictions of Einstein's equations that cannot be observed or detected, with the exception of wormholes (WH) \cite{Einstein:1935tc}. The high level of prediction of the EEq is astonishing, and many of them are surprising. 

In this work, we use the Einstein-Maxwell-Dilaton (EMD) theory to derive a new combined solution that generalises two previously investigated wormhole solutions: the asymptotically flat configuration \cite{DelAguila:2015isj} and the NUT spacetime incorporating a magnetic monopole and a scalar field \cite{Bixano:2025jwm}, thereby obtaining a new exact solution combination of the EEq which, in the framework of the Petrov classification, corresponds to a Petrov type I as shown in this paper, which includes electric and magnetic fields together with a scalar field that may be of dilaton type (i.e., real and coupled to the electromagnetic field) or of phantom type (i.e., featuring a negative kinetic term in the Lagrangian). This configuration is further separated into a super-extreme and a sub-extreme regime, which correspond, respectively, to a wormhole and to a new exact black hole solution (valid for any selected gravitational theory, coupled to a phantom/dilaton-type scalar field, and accommodating both asymptotically flat and NUT-type spacetime geometries).

In Einstein-Maxwell theory, the overloaded Kerr-Newman-NUT sector can be interpreted as a geodesically complete, traversable wormhole. In this case, the effective violation of the zero-energy condition is linked to Misner-Dirac strings (distributional sources) rather than a regular Maxwell field, and the conserved quantities (Komar-type charges) can be separated accordingly \cite{Clement:2022pjr}.
In Einstein-Maxwell-dilaton (EMD) theory, exact solutions of traversable wormholes have also been constructed and analyzed, including the appearance of a ring singularity that is considered inaccessible to causal geodesics, along with a discussion of the asymptotic properties in different branches \cite{Bixano:2025jwm}
On the other hand, numerical families of nut-shaped wormholes arise in scalar-tensor models of greater curvature (e.g., Gauss-Bonnet and Chern-Simons couplings), where the NUT loading implies a Misner string (hence the asymptotic non-planarity in the usual sense) and the appearance of a critical polar angle beyond which closed time-curves occur \cite{Ibadov:2020ajr}.
These works provide a useful basis for determining which physical properties NUT-supported wormholes share (asymptotic ALFs, Misner string effects, CTC sectors) and which are specific to the field content and construction method of the solutions considered here.

In this work, we clarify the mechanism of the Wormhole Cosmic Censorship Conjecture (WCCC) more effectively. Specifically, although a ring singularity exists, the wormhole throat hides causal spacetime pathologies. In particular, by examining the subextreme regime, we have shown that the event horizon similarly hides both causal pathologies and violations of energetic conditions, thus satisfying the Cosmic Censorship Conjecture (CCC). In this way, we provide a clearer characterization of the solution's regularity.
If we place our solution within the set of known configurations, we obtain a purely electromagnetic compact object whose Komar mass at infinity vanishes. This follows directly from the fact that the metric function \(f\) is constant and, therefore, there is no gravitational source. In other words, it is a purely electromagnetic object, endowed with a scalar field, that produces a rotating black hole or wormhole in spacetime. This becomes particularly evident when examined using invariant charges. An interesting peculiarity arises when the NUT parameter \(\tau_0 L_{\pm}/2\) is set to zero. We observe that the electric and magnetic flux charges are exactly equal and both depend on \(\tau_0\). Consequently, if this constant \(\tau_0\) is turned off, we obtain a configuration whose flux charge at infinity vanishes. Similarly, the Komar mass is zero, and since the NUT parameter is also zero, the resulting solution is a rotating black hole or wormhole with no charge at infinity and no mass, asymptotically flat, and yet with angular momentum \(J_\infty \neq 0\). This is a very remarkable feature.

We present them using two parameters: the parameter $\epsilon_0$, which takes the value 1 for the dilaton field and $-1$ for the phantom field, and the parameter $\alpha_0$ to distinguish them from different theories. In \cite{Minazzoli:2025nbi}, different values for the parameter $\alpha_0$ are used to derive solutions that correspond to examples of compact objects. The exact solution is valid for almost all values of these parameters, provided they satisfy a constraint equation. We have shown that this combination of solutions corresponds to a NUT configuration, and we demonstrate that the parameter $L_{\pm}\,\tau_0/2$ serves as the NUT parameter by determining the NUT charge and analyzing the asymptotic behavior of both the metric functions and the 4-potential. In addition, we confirm this identification through an independent method based on the Maxwell scalars constructed from the Newman–Penrose null tetrad, again by analyzing their asymptotic behavior.

We have analyzed the event horizons, closed time-like curves (CTCs), and the wormhole throat, and conclude that the super-extreme case, described in oblate spheroidal coordinates, corresponds to a wormhole (WH), whereas the sub-extreme case, also described in oblate spheroidal coordinates, corresponds to a black hole (BH). In the wormhole case, the solution features a ring singularity enclosed by the throat, while in the black hole case, we find the same type of ring singularity, along with two additional superficial singularities concealed behind the event horizon. Nevertheless, as in our previous WH solutions, we demonstrate here that geodesics in the WH, can never reach these singularities, they would require an infinite amount of time to get arbitrarily close to them. We call this phenomenon Wormhole Cosmic Censorship \cite{axioms14110831,Matos:2012gj,DelAguila:2018gni}, it has been observed in several exact WH solutions of the EEq \cite{Matos:2000ai,Matos:2000za,Matos:2005uh,Matos:2009rp,Matos:2010pcd}. Moreover, these solutions fulfill the energy conditions for $\epsilon_0 = 1$, and we identify the region in which they are traversable for WH or, in the black hole case, the exterior region beyond the event horizon. 
Using a diagonal tetrad formalism, we analyzed the tidal forces and projected various null geodesics onto the tidal-force surface in order to study the behavior of the corresponding trajectories. In an analogous manner, we computed and represented the vectorial electromagnetic field and again projected the same set of null geodesics onto this field configuration. From these constructions, it is evident that the geodesics tend to avoid regions of higher tidal-force magnitude and stronger electromagnetic field intensity. 

Furthermore, we carried out a detailed investigation of the geodesic structure associated with both wormhole and black hole spacetimes, employing spheroidal and Cartesian coordinate systems. This allowed us to examine in greater depth the geometric structure of the compact objects and the corresponding patterns of geodesic motion.

Finally, we show different possibilities for the size of these solutions where they can be realistic. This raises the possibility that some cosmic objects could be WH rather than other compact objects, provided that the dilaton interaction is possible.

In this instance, we start with a Lagrangian expressed in the International System of Units (SI), which has energy density units

{\setlength{\abovedisplayskip}{-5pt}
 \setlength{\abovedisplayshortskip}{-5pt}
\begin{multline}\label{LagrangianoTesisUnidades}
    \mathfrak{L}=\sqrt{-g}\bigg(-\frac{c^4}{8 \pi G}R +\frac{c^4}{8 \pi G}2\epsilon_0 (\nabla \phi)^2 \\ + \frac{1}{\mu_0}e^{-2 \alpha_0 \phi } F^2 \bigg),
\end{multline}
}
where $c$ is the speed of light, $G$ is the gravitational constant, and $\mu_0$ is the vacuum permeability. $\phi$ denotes the scalar field and is characterized as dimensionless. The parameter $\alpha_0$ specifies the theoretical framework employed and has the potential to adopt the values
\begin{equation}\label{Alpha0}
    \alpha_0^2 = \begin{cases} 
    1/12 & \Rightarrow \mathfrak{L} \text{ corresponds to}  \\ & \text{Entangled Relativity lagrangian}  \\
    3 & \Rightarrow \mathfrak{L} \text{ corresponds to}  \\ & \text{Kaluza-Klein lagrangian}  \\
    1 & \Rightarrow \mathfrak{L} \text{ corresponds to Low-Energy}  \\ & \text{Super Strings lagrangian }\\ 
    0 & \Rightarrow \mathfrak{L} \text{ corresponds to}  \\ & \text{Eintein-Maxwell lagrangian}
    \end{cases}
\end{equation}
 and $\epsilon_0$ take only two values, $1$ if $\phi$ is a dilatonic scalar field and $-1$ if it is a phantom scalar field. $F_{\mu \nu}$ is the Faraday tensor that has units of teslas, and $R$ is the Ricci scalar with length$^{-2}$ units.

Defining an important constant that provides the correct units for the 4-potential
{\small
\begin{equation}\label{ConstanteSgima}
    \sigma_0 \equiv \frac{8\pi G}{\mu_0 c^4} \qquad \text{implies} \quad \frac{1}{\sqrt{\sigma_0}}\thickapprox 2.46 \times 10^{18}   [\text{Meters*Teslas}],
\end{equation}
}
The corresponding field equations of the Lagrangian (\ref{LagrangianoTesisUnidades}) are:
{\setlength{\abovedisplayskip}{-6pt}
 \setlength{\abovedisplayshortskip}{-6pt}
\begin{subequations}\label{EcuacionesDeCampoOriginales}
\begin{equation}\label{Eq:Campo1}
    \nabla_\mu \left( e^{-2\alpha_0 \phi} F^{\mu \nu} \right)=0,
\end{equation}
\begin{equation}\label{Eq:Campo2}
    \epsilon_0 \nabla^2 \phi+\frac{\alpha_0}{2} \sigma_0 \left( e^{-2\alpha_0 \phi} F^{2} \right)=0,
\end{equation}
\begin{multline}\label{Eq:Campo3}
    R_{\mu \nu}=2\epsilon_0 \nabla_\mu \phi \nabla_\nu \phi \\ + 2 \sigma_0 e^{-2\alpha_0 \phi} \left( F_{\mu \sigma} \tensor{F}{_\nu}{^\sigma} -\frac{1}{4} g_{\mu \nu } F^2 \right),
\end{multline}
\end{subequations}
}
where $\nabla^2= \nabla^\mu \nabla_\mu$. The above equations are, in general, very difficult to solve. There is a way to solve them for the axisymmetric stationary case given in \cite{Matos:2010pcd}.
\section{Previous tools.}

Cylindrical or Weyl-coordinates $(\rho,z)$ will serve as a fundamental tool for performing physical analyzes. However, for mathematical processes, the spheroidal coordinates $(x,y)$ are more advantageous; the relationship between them is
\begin{equation}\label{RelacionCilindricasEsferoidales}
    \rho^2 =L_{\pm}^2(x^2\pm 1)(1-y^2), \qquad 
    z=L_{\pm}xy,
\end{equation}
where $\rho \in [0,\infty)$ ,$\{ z ,x \}\in \mathbb{R}$, and $y \in [-1,1]$, and $(L_{\pm}) \geq 0$. 

We are considering two scenarios, the first related to the sub-extreme condition (S-E:$-$):
\begin{align}\label{Sub-Extreme}
|\mathcal{M}_\infty|^{2}&>a^2+Q_{L}^{2}+H_{L}^{2} ,\\
|\mathcal{M}_\infty|^{2}&=L_{-}^2+a^2+Q_{L}^{2}+H_{L}^{2},
\end{align}
where $a = J_\infty / |\mathcal{M}_\infty|$ denotes the moment per unit effective mass, with $\mathcal{M}_\infty = l_1 + i N_\infty$, $l_1$ is a parameter that has the dimensions of length, and $Q_L, H_L$ refer to the electric and magnetic geometric charges. The second case corresponds to the super-extreme configuration (SU-E:$+$)
\begin{align}\label{Super-Extreme}
|\mathcal{M}_\infty|^{2}&<a^2+Q_{L}^{2}+H_{L}^{2} ,\\
L_{+}^2+|\mathcal{M}_\infty|^{2}&=a^2+Q_{L}^{2}+H_{L}^{2}.
\end{align}
In both cases, the solution is consistent because the expressions in the metric change to $L^2(x^2+y^2) \rightarrow L^2(x^2-y^2)$ and $L^2(x^2+1) \rightarrow L^2(x^2-1)$.
This criterion is inspired by the horizon/root structure of Kerr–Newman–Taub–NUT spacetimes, where the key radial function is a quadratic polynomial whose discriminant specifies whether the candidate horizon radii are real (sub-extreme), coincident (extreme), or complex (super-extreme). In this framework, the NUT parameter enters the discriminant on equal footing with the combination \(l_1{}^{2}+N^{2}\), while rotation and electric/magnetic charges appear via \(a^{2}+Q_L^{2}+H_L^{2}\). A similar discriminant-based classification has been used in closely related studies; see references \cite{Clement:2022pjr,Clement:2015aka,Paganini:2017qfo,Pradhan:2014zia}.
Ultimately, the Boyer-Lindquist coordinates $(r,\theta)$ correspond to the earlier coordinates as
{\setlength{\abovedisplayskip}{0pt}
 \setlength{\abovedisplayshortskip}{0pt}
\begin{equation}\label{BoyerLindquistSpheroidalCordiantes}
    (L_{\pm}) x=r-l_1, \qquad y=\cos{\theta},
\end{equation}
}
where $r \in (-\infty,-l_1]\cup[l_1,\infty)$, $\theta\in [0,\pi]$.

An important point is that, to characterize the size of the wormhole, we directly identify the throat radius $l_1$, equal to half of Schwarzschild radius, i.e. $l_1 = r_s/2$, thus we fix the size of the wormhole and, from this relationship, derive the nature of the object. For example, for a wormhole the size of the sun, we would use $l_1 \approx 1.5\,$ km. For a black hole, the parameter $l_1$ is related to the Schwarzschild radius.
In this work, $l_1$ generally differs from the geometric mass, i.e. in general $l_1 \neq M_{\text{geo}} = G M_{\text{SI}}/c^2$, with $M_{\text{SI}}$ the mass in SI units.

The variables $\{\rho, z, L_{\pm}\}$ have units of length. The variables $\{x,y\}$ are dimensionless.

To express the parameters in SI, we must consider the expressions
\begin{subequations}\label{TodasLasConstantesConUnidades}
\begin{align}
    [\, |\mathcal{M}_\infty| \equiv \frac{c^2}{G} |\mathcal{M}|\, ] \qquad &= \quad\text{Length},\\
    [J_\infty \equiv \frac{GJ}{c^3}] \qquad &= \quad\text{Length}^2,\\
    [a\equiv \frac{J_\infty}{|\mathcal{M}|}]\qquad &= \quad\text{Length} ,\\
    [Q_L \equiv \sqrt{\sigma_0} \, Q_{\infty} ] \qquad &= \quad\text{Length} ,\\
    [H_L \equiv \sqrt{\sigma_0} \, H_{\infty} ] \qquad &= \quad\text{Length} ,  
\end{align}
\end{subequations}
where $J$ possesses units of angular momentum, $|\mathcal{M}|$ corresponds to units of mass, and $\{ Q_\infty, H_\infty \}$ are characterized by the units of lenght$^2*$tesla, respectively.

We begin with a space-time characterized by two commutative symmetries, meaning there are two Killing vectors, $\partial _\varphi$ and $\partial_t$. These vectors permit the formulation of the Weyl ansatz metric in spheroidal coordinates $(x, y)$, either Oblates ($+$) or Prolates ($-$), denoted as:
\begin{multline}\label{ds sp}
    ds^2 = -f\left( cdt-\omega d \varphi \right)^2
     + \frac{(L_{\pm})^2}{f} \bigg( (x^2\pm1)(1-y^2) d\varphi^2 \\
     +(x^2\pm y^2) e^{2k} \left\{ \frac{dx^2}{x^2\pm 1} +\frac{dy^2}{1-y^2} \right\} \bigg).
\end{multline}

where $\{f,\omega,\kappa \}$ are the metric functions that depend on the coordinates $(\rho,z)$, $\{f,\kappa \}$ are dimensionless, and $\omega$ has units of length.

An important point is that, throughout the paper, the upper sign will correspond to the super-extreme case, which will be associated with oblate coordinates and gives rise to the wormhole satisfying \eqref{Super-Extreme}, whereas the lower sign in our expressions will always correspond to the sub-extreme case, which will be associated with prolate coordinates and gives rise to the black hole, satisfying \eqref{Sub-Extreme}. In the opposite case, when there are not two symbols, it will be clarified which compact object is being referred to.

Using the symmetries of space-time, the electromagnetic 4-Potential is reduced to
\begin{equation}\label{4Potencial}
    A_{\mu}=\bigg[ A_t(\rho,z),0,0,L_{\pm}\,A_\varphi (\rho,z) \bigg] .
\end{equation}
%
%


Next, we solve the equations (\ref{EcuacionesDeCampoOriginales}) using the method proposed in \cite{Matos:2000ai}. To do this, the method defines the potentials.
{\setlength{\abovedisplayskip}{-6.8pt}
 \setlength{\abovedisplayshortskip}{-6.8pt}
\begin{subequations}\label{DefinicionPotenciales}

    \begin{equation}
        \Tilde{D}\chi=\sigma_0\frac{2f\kappa^2}{\rho} L(\frac{\omega}{L} DA_t +DA_\varphi),
    \end{equation}
    \begin{equation}
        \Tilde{D} \epsilon = \frac{f^2}{\rho} D\omega + \psi \Tilde{D}\chi,
    \end{equation}
    \begin{equation}
        \psi=2A_t,
    \end{equation}
\end{subequations} 
}

where the differential operators $D, \Tilde{D}$ are 
\begin{equation}
    D=
    \begin{bmatrix}
    \begin{array}{c}
    \partial_\rho \\
    \partial_z
    \end{array}
    \end{bmatrix},
    \qquad
    \Tilde{D}=
    \begin{bmatrix}
    \begin{array}{c}
    \partial_z \\
    -\partial_\rho
    \end{array}
    \end{bmatrix}.
\end{equation}
To simplify the notation we define the vector potential $Y^A$
\begin{equation}\label{Potenciales}
    [Y^A]=\begin{bmatrix}
        f \\
        \epsilon \\
        \psi \\
        \chi \\
        \kappa
    \end{bmatrix} \quad = \quad \begin{bmatrix}
        \text{dimensionless} \\
        \text{dimensionless} \\
        \text{length * tesla} \\
        (\text{length * tesla})^{-1} \\
        \text{dimensionless}
    \end{bmatrix},
\end{equation}
which are respectively the gravitational, rotational, electric, magnetic and scalar potentials.

Substituting the metric $ds^2 = -f\left( cdt-\omega d \varphi \right)^2+f^{-1}e^{2k}(d\rho ^2+dz^2)+f^{-1}\rho^2d\varphi^2$, the 4-potential (\ref{4Potencial}), and the definitions (\ref{DefinicionPotenciales}) into the field equations (\ref{EcuacionesDeCampoOriginales}), these reduce to
\begin{subequations}\label{EcuacionesDeCampoV2}
\begin{center}
    \text{\textbf{Klein-Gordon equation}}
    \begin{equation}
        D(\rho D\kappa ) - \frac{\rho }{\kappa} D\kappa^2+\frac{\rho \kappa^3 \alpha_0^2}{4 f \epsilon_0} \sigma_0 \left( D\psi^2-\frac{1}{\kappa^4 (\sigma_0)^2} D\chi^2 \right) =0 \label{eq:KleinGordonV2} ,
    \end{equation}
    \text{\textbf{Maxwell equations}}
    \begin{multline}\label{Eq:Maxwell1V2}
        D (\rho D\psi)+\rho D\psi \left( \frac{2D\kappa}{\kappa} -\frac{Df}{f} \right) \\ -\frac{\rho}{f\kappa^2 \sigma_0} (D\epsilon -\psi D\chi) D\chi =0 ,
    \end{multline}
    \begin{multline}\label{Eq:Maxwell2V2}
        D (\rho D\chi)-\rho D\chi \left( \frac{2D\kappa}{\kappa} +\frac{Df}{f} \right) \\ +\frac{\rho \kappa^2}{f} \sigma_0(D\epsilon -\psi D\chi) D\psi =0, 
    \end{multline}  
    \text{\textbf{Einstein equations}}
    \begin{multline}\label{Eq:Einstein1V2}
        D (\rho Df)+\frac{\rho}{f}\left( (D\epsilon-\psi D\chi )^2 -Df^2 \right) \\ -\frac{\rho \kappa^2}{2} \sigma_0\left( D\psi^2 +\frac{1}{\kappa^4 (\sigma_0)^2} D\chi^2 \right) = 0 ,
    \end{multline}
    \begin{equation}\label{Eq:Einstein2V2}
        D (\rho (D\epsilon-\psi D\chi)) - \frac{2 \rho }{f} (D\epsilon-\psi D\chi) Df =0. 
    \end{equation}  
\end{center}
\end{subequations}
The next step in the method is to use the ansatz $Y^A=Y^A(\lambda)$, where $\lambda=\lambda(x,y)$, satisfies plane Laplace's equation for spheroidal Oblates($+$)/Prolates($-$) coordinates
\begin{equation}\label{Eq:LaplaceEnEsferoidales}
        \partial_x \{ (x^2\pm 1)\partial_x \lambda \}+\partial_y \{(1-y^2) \partial_y \lambda \}=0.
\end{equation}
With this ansatz, the field equation is considerably reduced and can be solved in some way, see \cite{Matos:2000za} and \cite{Matos:2000ai}. Using the method described in the mentioned papers allows the derivation of the second class of solutions:
{\setlength{\abovedisplayskip}{0pt}
 \setlength{\abovedisplayshortskip}{0pt}
\begin{multline}\label{SegundaClaseSoluciones}
    f=f_0, \quad \kappa=\kappa_0 e^{\lambda}, \quad \psi= \frac{\sqrt{f_0}}{\sqrt{\sigma_0} \kappa_0} e^{-\lambda} +\psi_0, \\ \quad \chi = \sqrt{\sigma_0} \sqrt{f_0}\kappa_0 e^{\lambda} + \chi_0, \quad \epsilon= b_0,
\end{multline} 
}
where $\{f_0, \psi_0, \chi_0, \kappa_0, b_0 \}$ are integration constants.

From the definitions of potentials (\ref{DefinicionPotenciales}), we can integrate the components of the metric \footnote{The differential equation for $k$ arises from the integration of one of the components of the field equation \ref{Eq:Campo3}.}
{\setlength{\abovedisplayskip}{-5pt}
 \setlength{\abovedisplayshortskip}{-5pt}
\begin{subequations}\label{FuncionesMetricas-xy}
\begin{center}
    \begin{equation}\label{EcuacionesDiferencialesOmega}
        \begin{bmatrix}
            \partial_x \\
            \partial_y \\
        \end{bmatrix} 
        \omega= \frac{(L_{\pm})}{f^2}\Big( \epsilon_{,\lambda} -\psi \chi_{,\lambda} \Big)
        \begin{bmatrix}
            (1-y^2)\partial_y  \\
            -(x^2\pm 1)\partial_x  
        \end{bmatrix}\lambda ,
    \end{equation}
    \begin{multline}\label{EcuacionesDiferencialesA3}
        \begin{bmatrix}
            \partial_x \\
            \partial_y \\
        \end{bmatrix} 
        A_\varphi = \frac{1}{2 \sigma_0 f \kappa^2 }\Big( \chi_{,\lambda} \Big) 
        \begin{bmatrix}
            (1-y^2)\partial_y  \\
            -(x^2\pm  1)\partial_x  
        \end{bmatrix} \lambda
        \\ -\frac{\omega}{2(L_{\pm})} \Big( \psi_{,\lambda} \Big)
        \begin{bmatrix}
            \partial_x \\
            \partial_y \\
        \end{bmatrix} \lambda,
    \end{multline}
    \begin{multline}\label{EcuacionesDiferencialesk1}
        \partial_x k=k_{0}\frac{1-y^2}{x^2\pm  y^2} \Big\{-2y (x^2\pm 1)(\partial_x \lambda)(\partial_y \lambda) \\ +x\big[ (x^2\pm 1)(\partial_x \lambda)^2-(1-y^2) (\partial_y \lambda)^2  \big]\Big\}, 
    \end{multline}
    \begin{multline}\label{EcuacionesDiferencialesk2}
        \partial_y k=k_{0}\frac{x^2\pm 1}{x^2\pm y^2} \Big\{ 2x (1-y^2)(\partial_x \lambda)(\partial_y \lambda) \\ +y\big[ (x^2\pm 1)(\partial_x \lambda)^2-(1-y^2) (\partial_y \lambda)^2  \big] \Big\}, 
    \end{multline}
\end{center}
\end{subequations}
}
where $k_0$ is an integration constant, $F_{,\lambda}=\partial F / \partial \lambda$, for any function $F$, and for SU-E($+$)/S-E($-$) scenario.

While the equations (\ref{EcuacionesDeCampoV2}), (\ref{SegundaClaseSoluciones}) and (\ref{FuncionesMetricas-xy}) present similarities with those presented in previous studies, a distinctive feature of this article is the inclusion of units and the extension to the S-E scenario using prolate coordinates, which improves the capacity of the equations for a meaningful physical interpretation.

Two solutions taken from \cite{Matos:2000ai}, of equation (\ref{Eq:LaplaceEnEsferoidales}), are
\begin{subequations}\label{LambdaSoluciones}
   \begin{align}
       \lambda_5&=\lambda_0 \frac{x}{(x^2\pm y^2)} , \label{lambda5}\\
       \lambda_{6}&=\lambda_0 \frac{y}{(x^2\pm y^2)}, \label{lambda6}
   \end{align} 
\end{subequations}
where $\lambda_0$ is an integration constant and for SU-E($+$)/S-E($-$) configuration.

The SU-E solutions have already been analyzed in \cite{Bixano:2025jwm} for $\lambda_5$ and in \cite{Matos:2010pcd} for $\lambda_6$. However, here we will explore the S-E scenario, specifically when $|\mathcal{M}_\infty|^{2}>a^2+Q_L^{2}+P_L^{2} $, and we will show that each compact object is entirely distinct.

The functions metric and the 4-potential to $\lambda_5$ are
{\setlength{\abovedisplayskip}{-10pt}
 \setlength{\abovedisplayshortskip}{-10pt}
{\small
\begin{subequations}\label{SolucionLambda5}
\begin{align}
        f&=f_0=1,\\
        \omega &=-(L_{\pm})\frac{ \lambda_0}{f_0 } \bigg( \frac{y(x^2\pm 1)}{x^2\pm y^2} \bigg), \label{Omega Lambda5} \\
        A_\varphi &=-\frac{\sqrt{f_0}}{2 \kappa_0 \sqrt{\sigma_0}} \frac{\omega}{(L_{\pm})} e^{-\lambda_5}, \label{A3 Lambda5} \\
        A_t&=\frac{\sqrt{f_0}}{2\kappa_0 \sqrt{\sigma_0}} e^{-\lambda_5}, \\
        k_{\lambda_5} &= -k_{0} \lambda_0 ^2  \frac{(1-y^2)}{4(x^2\pm y^2)^4} \bigg(  -8x^2y^2(x^2\pm 1)  \notag \\  &+[x^2\pm y^2]^2\big[(1-y^2) +2(x^2\pm y^2)\big] \bigg) . \label{k Lambda5}
\end{align}
\end{subequations} }
}
%
On the other hand, for $\lambda_{6}$ is
{\small
\begin{subequations}\label{SolucionLambda6}
\begin{align}
        f&=f_0=1, \\
        \omega &=\frac{ (L_{\pm}) \lambda_0 }{f_0} \frac{x(1-y^{2})}{x^{2}\pm y^{2}}, \label{Omega Lambda6}\\
        A_\varphi &=-\frac{\sqrt{f_0}}{2 \kappa_0 \sqrt{\sigma_0}} \frac{\omega}{(L_{\pm})} e^{-\lambda_{6}}, \label{A3 Lambda6} \\
        A_t&=\frac{\sqrt{f_0}}{2\kappa_0 \sqrt{\sigma_0}} e^{-\lambda_{6}}, \\
        k_{\lambda_6} &= -k_{0} \lambda_0 ^2  \frac{(1-y^2)}{4(x^2\pm y^2)^4} \bigg(  8x^2y^2(x^2\pm 1) \notag \\
        &-[x^2\pm y^2]^2(1-y^2)  \bigg) . \label{k Lambda6} 
\end{align}
\end{subequations} }
%
Both regular solutions in the SU-E scenario represent a rotating electromagnetic WH with no gravitational potential.
\section{Combination solution}

Using the linearity of the equation (\ref{Eq:LaplaceEnEsferoidales}), we can obtain a new solution by combining two or more of these. Two interesting solutions are (\ref{SolucionLambda6}) and (\ref{SolucionLambda5}). The combined solution $\lambda_c$ reads 
\begin{equation}\label{Lambda5+Lambda6}
    \lambda_c=\frac{\lambda_0 y + \tau_0 x}{(x^2\pm y^2)}.
\end{equation}
The corresponding dilatonic scalar field takes the form
\begin{equation}\label{FormaDelPotencialSuma}
    \phi(x,y)=-\frac{\lambda_c}{\alpha_0}.
\end{equation}
whose asymptotic behavior is $\lim\limits_{x \rightarrow \pm \infty } \phi(x,y)= 0$ which implies that the scalar field remains bounded.

Thus, the metric functions of the solution (\ref{Lambda5+Lambda6}) obtained from the equations (\ref{FuncionesMetricas-xy}) are
\begin{subequations}\label{SolucionLambdaCombinada}
\begin{align}
        f&=f_0=1, \\
        \omega &=\frac{ (L_{\pm}) }{f_0} \bigg( \frac{\lambda_0 x(1-y^{2})-\tau_0 y (x^2\pm 1)}{x^{2}\pm y^{2}} \bigg), \label{Omega LambdaCombinada}\\
        A_\varphi &=-\frac{\sqrt{f_0}}{2 \kappa_0 \sqrt{\sigma_0}} \frac{\omega}{(L_{\pm})}e^{-\lambda_{c}} , \label{A3 LambdaCombinada} \\
        A_t&=\frac{\sqrt{f_0}}{2\kappa_0 \sqrt{\sigma_0}} \bigg( e^{-\lambda_{c}} -1\bigg) \label{A0 LambdaCombinada}\\
        k_{c} &= k_{\lambda5}+k_{\lambda6} \notag \\
         & -k_{0}\frac{8xy(1-y^2)(x^2\pm 1)(x^2-y^2)\lambda_0 \tau_0}{4(x^2\pm y^2)^4}  \label{k LambdaCombinada},
\end{align}
\end{subequations}
It is important to emphasize that, although \(\lambda_c\) is a linear combination of \(\lambda_5\) and \(\lambda_6\), this does not imply that the metric functions or the four-potential are themselves linear. The linearity holds only within the auxiliary two-dimensional subspace introduced to simplify the structure of the five-dimensional potential space \(Y^A\). Consequently, the linearity of \(\lambda\) in this reduced subspace does not translate into linear behavior in the full potential space, and therefore the corresponding solution in physical spacetime coordinates \((ct,\rho,z,\varphi)\) is inherently non-linear. 

In an analogous way, we have obtained a new exact black hole solution, valid for any underlying gravitational theory of interest, coupled to a scalar field of phantom/dilaton type, admitting both asymptotically flat and NUT-type configurations.

Note that the asymptotic behavior ($ r\rightarrow \pm \infty \Rightarrow x\rightarrow \pm \infty $) of the metric functions is as follows
{\small
\begin{subequations}\label{ComportamientoAsimptoticoOmega}
\setlength{\jot}{0pt}  
    \begin{align}
        \lim\limits_{x \rightarrow \pm \infty } \omega(x,y) &= - \tau_0 (L_{\pm}) y, \label{OmegaxInfinito} \\
        \omega(0,y) &= - \frac{\tau_0 (L_{\pm})}{y},\label{OmegaxCero} \\
        \omega(x,0) &= \frac{(L_{\pm}) \lambda_0}{x},\label{omegay0} \\
        \omega(x,1) &=-(L_{\pm}) \tau_0,\label{omegay1}
    \end{align}
\end{subequations}
}
{\setlength{\abovedisplayskip}{-10pt}
 \setlength{\abovedisplayshortskip}{-10pt}
\begin{subequations}\label{ComportamientoAsimptoticoK}
\setlength{\jot}{0pt}  
    \begin{align}
        \text{For} \quad x \gg 1 \,\,\,\,\,\, \text{we have}\,\,\,\,\,\,  k_c \approx - k_0 \tau_0^2 \frac{(1-y^2)}{2x^2}, \label{kxInfinito} \\
        k_c(0,y)= \frac{k_0(1-y^2)}{4y^4} \big\{ \lambda_0^2 (1-y^2) -\tau_0^2 (1+y^2) \big\}, \label{kxCero} \\
        k_c (x,0)= k_0\frac{(\lambda_0^2-\tau_0^2(2x^2+1))}{
        4x^4}, \label{ky0} \\
        k_c (x,1)=0. \label{ky1}
    \end{align}
\end{subequations}
}
%
In the same way as in \cite{Bixano:2025jwm}, and \cite{DelAguila:2015isj}, we substitute the solution (\ref{SolucionLambdaCombinada}) associated with $\lambda_c$, and the scalar field (\ref{FormaDelPotencialSuma}), in (\ref{EcuacionesDeCampoOriginales}) to obtain a constraint on the free parameters of the solution, this is

{\setlength{\abovedisplayskip}{-10pt}
 \setlength{\abovedisplayshortskip}{-10pt}
\begin{equation}\label{EcuacionDeVerificacion}
    \alpha_0^2(4k_0+1)-4\epsilon_0=0 .
\end{equation}
}
In Table \ref{TablaValoresk} we present the allowed values for the $k_0$ parameter (see also \cite{DelAguila:2015isj} and \cite{Bixano:2025jwm}). 

\begin{table}[b]
\caption{\label{TablaValoresk}%
 Values of $k_0$ for SU-E/S-E scenarios }
\begin{ruledtabular}
\begin{tabular}{ccc}
\textbf{$\alpha_0^2$} &  
  \textbf{Dilatonic field ($\epsilon_0=1$)} & \textbf{Phantom field ($\epsilon_0=-1$)} \\
\colrule
1/12 & $47/4$ & $-49/4$ \\
1 & $3/4$ & $-5/4$ \\
3 & $1/12$ & $-7/12$ \\
4 & 0 & $-1/2$ 
\end{tabular}
\end{ruledtabular}
\end{table}

\section{Ring singularity and asymptotical flatness}

To study the metric singularities of space-time, we can use two important scalars: the Ricci scalar $R=\tensor{g}{^{\alpha \beta}} \tensor{R}{_{\alpha \beta}}=\tensor{g}{^{\alpha \beta}} \tensor{R}{^\sigma}{_{\alpha \sigma \beta}}$ and the Kretschmann scalar $KN=\tensor{R}{^{ \mu \nu \alpha \beta}}\tensor{R}{}{_{ \mu \nu \alpha \beta}}$. The invariants $R$ and $KN$ associated to ($\lambda_c$) are
\begin{subequations}\label{Invariantes R KN}
\begin{widetext}
    \begin{equation}
        R=\frac{(4k_0+1) e^{-2k_c}}{2L_\pm^2 (x^2\pm y^2)^4} \bigg\{ \lambda_0^2 \Big( \pm y^2(1-y^2) +x^2(3y^2+1)  \Big) +\tau_0^2 \Big( y^2 \pm x^2 +x^2(x^2 \mp 3y^2) \Big) +4\lambda_0 \tau_0 x y (x^2\mp y^2) \bigg\}, \label{Ricci para lambdac}
    \end{equation}
\end{widetext}
    \begin{align}
        KN=\frac{F_{5}(x,y)}{4L_\pm^{4}(x^2\pm y^2)^{12}} e^{-4k(x,y)}, \label{KN para lambdac}
    \end{align}
\end{subequations}
where, $F_{5}(x,y)$ is a polynomial of degree less than $(x^2\pm y^2)^{12}$.

Using (\ref{kxInfinito}), we obtain $\lim\limits_{x \rightarrow \pm \infty } e^{-k_c}=1$ regardless of the sign of $k_0$ and for both the SU-E (Upper sign) and S-E (Lower sign) cases. Consequently, the asymptotic form of (\ref{Ricci para lambdac}) becomes
{\setlength{\abovedisplayskip}{-2pt}
 \setlength{\abovedisplayshortskip}{-2pt}
\begin{equation}\label{RicciXInfinito}
     \text{for} \quad x \gg 1 \,\,\,\,\,\,\text{we have}\,\,\,\,\,\, R\approx (4k_0+1)\frac{\tau_0^2}{2L_\pm^2 x^4},
\end{equation}
}
therefore, $R$ and $KN$ tend to zero for any sign of $(4k_0+1)$, in other words, the matter fields become asymptotically trace-free as $x\rightarrow \pm \infty$ for both the dilatonic and phantom scalar fields.
On the other hand, (\ref{Invariantes R KN}) reveals the presence of a ring singularity at $x=0$ and $y=0$, which corresponds to $r=l_1$ and $\theta=\pi/2$ in Lewis–Papapetrou coordinates, or to $\rho=L_{\pm}$ and $z=0$ in Weyl coordinates, \textit{within the SU-E case}. In contrast, for the S-E configuration, we obtain the same ring singularity, but in this situation two additional surface singularities arise, located at $x=\pm y$ or, equivalently, at $r=l_1\pm (L_{\pm})\cos{\theta}$.
%
\subsection{Asymptotic physical interpretation}

In the asymptotic region $r\to \pm \infty$ the electromagnetic potential tends to
\begin{equation}\label{Comportamiento Asimptotic 4Potential}
A \simeq -\frac{L_{\pm }\, \tau_0}{2r}dt+ \frac{L_{\pm }\,\tau_0}{2}\cos\theta d\varphi\,,
\end{equation}
then the Maxwell field components behave at leading order as $F_{tr} \sim (L_{\pm}\tau_0)/(2r^2)$,
$F_{t \theta } \sim (L_{\pm}^2\lambda_0 \sin{\theta})/(2r^2)$,
$F_{ \theta \varphi} \sim L_{\pm}(\tau_0\sin{\theta})/2$ and
$F_{\varphi r} \sim L_{\pm}^2(\tau_0^2\cos\theta + \lambda_0\sin^2\theta)/(2r^2)$. At the same time, the first invariant $F_{\mu\nu}F^{\mu\nu}$ is identically zero for all $x$. 

In the following section, \ref{Invariant conserved charges (Komar / dual Komar / EM fluxes)}, we calculate the electric and magnetic flux charge at infinity ($H_\infty, Q_\infty$), we can see that $Q_\infty = H_\infty$. Consequently, the solution represents a Dionic configuration at infinity, endowed with non-annulling electric and magnetic monopole charges (see, for example, \cite{Ramirez-Valdez:2023gkh, Herdeiro:2024yqa, Chen:2012pt}).

Furthermore, since asymptotically we have \(F_{\theta\varphi}\propto \sin\theta\), the orthonormal radial component of the magnetic field behaves as
\(B^{\hat r}\sim H/r^{2}\) and is regular on the entire sphere \(S^{2}_{\infty}\); see the following section \ref{Invariant conserved charges (Komar / dual Komar / EM fluxes)}. Therefore, the electromagnetic field at infinity is that of a standard magnetic monopole and does not require any additional Dirac string-like singularity in the norm sector. Conversely, a string-like structure can arise in the gravitational sector: the asymptotic drag potential contains a NUT-like contribution. From \eqref{Omega LambdaCombinada} we find, for \(x\to\infty\) (equivalently \(r\simeq L_\pm x\), \(y=\cos\theta\)),
\begin{equation}\label{eq:gtphi_asymptotic}
g_{t\varphi}=f\,\omega
=
-\,L_\pm \tau_0\cos\theta
+\frac{L_\pm^{\,2}\lambda_0}{r}\sin^{2}\theta
+\mathcal O(r^{-2}).
\end{equation}
The principal term \(\cos\theta\) is the characteristic ALF/Misner string (gravitomagnetic) contribution of a NUT spacetime, while the subprincipal tail \(\sin^{2}\theta/r\) is the standard rotational frame-drag term (as in Kerr/Kerr-Newman). This explicitly shows why a rigorous distinction between angular momentum and NUT charge is required, even though both enter \(g_{t\varphi}\). They correspond to different angular sectors and drops; to see this, we compute \(J\) and \(N\) from invariant conserved charges (Komar for \(J\) and a dual/Komar-like definition for \(N\)) rather than identifying them directly from \(\omega\) alone \cite{Komar:1958wp,Taub:1950ez,Newman:1963yy,Gibbons:1979xm,Nedkova:2011hx,Awad:2022jgn} (see the section \ref{Invariant conserved charges (Komar / dual Komar / EM fluxes)}). In this framework, the parameter \(\tau_{0}\) controls the gravitomagnetic sector (NUT/Misner-string) and simultaneously fixes the magnetic flux charge in the electromagnetic sector, while \(\lambda_{0}\) controls the genuine rotational sector through the contribution of \(\sin^{2}\theta/r\) in \eqref{eq:gtphi_asymptotic}.

Finally, we note that the electromagnetic invariant \(F_{\mu\nu}F^{\mu\nu}=0\) holds identically for all \((x,y)\). In particular, this
does not imply that the electromagnetic field vanishes; rather, it indicates a null-type field configuration (invariantly) while the
conserved flux charges \(H\) and \(Q\) remain non-trivial.

Consequently, even though the spacetime is asymptotically Ricci-flat and curvature invariants such as the Ricci scalar and the Kretschmann scalar vanish in the limit $r \to \infty$, the existence of a nontrivial gravitomagnetic potential and of Misner strings implies that the metric does \emph{not} approach the Minkowski form at either spatial or null infinity. The spacetime is therefore only asymptotically locally flat (ALF) in the Taub–NUT sense, and it fails to be asymptotically flat in the standard Bondi–Sachs sense \cite{Bondi:1962px,Sachs:1961zz,Penrose:1962ij}.

In addition, choosing $\tau_0 = 0$ eliminates the NUT-like term and produces an asymptotically flat (AF) geometry ($g_{\mu \nu} \xrightarrow[r\to \pm\infty]{} \eta_{\mu \nu}$, where $\eta_{\mu \nu}$ denotes the Minkowski metric). In other words, the combination \eqref{SolucionLambdaCombinada} is not merely the linear superposition of \eqref{A3 Lambda5} and \eqref{A3 Lambda6}.

\section{Invariant conserved charges (Komar / dual Komar / EM fluxes)}\label{Invariant conserved charges (Komar / dual Komar / EM fluxes)}
 
In this section, we will calculate the invariant conserved charges using the results from Appendix \ref{ApnediceCargasInv}.
The mass $M$ is calculated from the Komar integral of the stationary Killing vector $\partial_t$, following Komar's original construction and modern treatments (including gauge fields) and, in ALF/Misner string geometries (NUT), works that explicitly address this, \cite{Komar:1958wp,Nedkova:2011hx,Clement:2015aka,Clement:2022pjr}. The angular momentum $J$ was obtained as the Komar charge of the axial Killing vector $\partial_\varphi$, using formulations that isolate $J$ from the contributions of Misner strings and comparing it with analyses of rotating wormholes with NUTs, \cite{Komar:1958wp,Nedkova:2011hx,Clement:2022pjr,BallonBordo:2019vrn}. The NUT charge $N$ was defined as the \emph{gravitomagnetic charge} (Komar dual) of the stationary Killing vector and, equivalently, via the NUT potential and its Smarr/first law type relations on ALF geometries, using invariant constructions of $N$ that clarify its physical meaning and the role of Misner strings \cite{Nedkova:2011hx,BallonBordo:2019vrn,Misner:1963fr,Manko:2005nm}.

Let the Killing 1-forms
\begin{align}
({}_tK)^\flat&=g_{\mu\nu}({}_tK)^\nu d x^\mu=g_{t\mu}d x^\mu, \notag\\
&= g_{tt}\,dt +g_{t \varphi} \,d\varphi \notag \\
&= -f\,d t+f\omega\,d\varphi \label{1FormaKillingt}\\
 \notag \\
({}_\varphi K)^\flat&=g_{\mu\nu}({}_\varphi K)^\nu d x^\mu=g_{\varphi\mu}d x^\mu \notag \\
&= g_{\varphi t}\,dt +g_{\varphi \varphi} \,d\varphi \notag \\
&=f\omega\,dt+\left(\frac{\rho^2}{f}-f\omega^2\right)d\varphi \label{1FormaKillingphi}.
\end{align}
Therefore, calculating the exterior derivative of \eqref{1FormaKillingt}
\begin{align*}
    d(({}_tK)^\flat)&=-(\partial_x f)\,dx\wedge d t-(\partial_y f)\,d y\wedge d t \notag \\
    &+(\partial_x(f\omega))\,d x\wedge d\varphi+(\partial_y(f\omega))\, d y\wedge d\varphi,
\end{align*}
where in $S_x$ we have $d t=0$ and $d x=0$, so only the term $d y\wedge d\varphi$ survives
\[
\Big[d(({}_tK)^\flat) \Big]\Bigg|_{S_x}=(\partial_y(f\omega))\, d y\wedge d\varphi.
\]
Thus, the NUT charge is determined by the corresponding Komar dual integral, obtaining
{\small
\begin{align}\label{eq:NUT_dualKomar}
N(x)&= -\frac{1}{8\pi}\int_{S_x} d(({}_tK)^\flat)= -\frac{1}{8\pi}\int_0^{2\pi}d \varphi\int_{-1}^{1} d y\;\partial_y(f\omega) \notag \\
&=-\frac{f \omega}{4}\Bigg|_{y=-1}^{y=1}.
\end{align}
}
Therefore, the NUT charge at that infinity is
\begin{equation}\label{eq:NUT_dualKomarInfinito}
    N_\infty=\lim_{x\to\infty}N(x)=\frac{\tau_0\, L_{\pm}}{2}.
\end{equation}

Now we calculate $M(x)$. We set $H=d (({}_tK)^\flat)$ in \eqref{eq:HodgeHMaestra}, we obtain
{\small
\begin{align*}
\Big[*d(({}_tK)^\flat)\Big]_{y\varphi}
&=L_{\pm}\, (x^2\pm1)\Big(g^{tt}\Big\{ d(({}_tK)^\flat)\Big\}_{tx}\notag \\ 
&\quad +g^{t\varphi}\Big\{ d(({}_tK)^\flat)\Big\}_{\varphi x}\Big)\nonumber\\
&=L_{\pm}\, (x^2\pm1)\Big(g^{tt}\partial_x f - g^{t\varphi}\partial_x(f\omega)\Big) \notag \\
&=-L_{\pm}\, (x^2\pm1) \left[\partial_x(\ln f)+\frac{f^2\omega}{\rho^2}\,\partial_x\omega\right].
\end{align*}
}
The Komar mass is defined as
{\small
\begin{align}\label{eq:Masa_Komar}
    M(x)&= \frac{-1}{8\pi}\int_{S} *d(({}_tK)^\flat)=\frac{-1}{8\pi}\int_0^{2\pi} d\varphi \int_{-1}^{1} dy \Big[*d(({}_tK)^\flat)\Big]_{y\varphi} \notag \\
    &=-\frac{L \left(\lambda _0^2+\tau _0^2\right)}{8 \left(x^4+x^2\right)}\left(x(1-x^2)+\left(x^2+1\right)^2 \cot^{-1}(x)\right).
\end{align}
}
Therefore, the (Komar) mass associated with that asymptotic region is
\begin{equation}\label{eq:Masa_KomarInfinito}
    M_{\infty}=\lim_{x\to\infty}M(x)=0.
\end{equation}

Substituting \eqref{1FormaKillingphi} into \eqref{eq:HodgeHMaestra}, we obtain
{\small
\begin{align*}
    \Big[*d(({}_\varphi K)^\flat)\Big]_{y\varphi}&=-L_{\pm}\, (x^2\pm1) \Big(g^{tt}\partial_x(f\omega)+g^{t\varphi}\partial_x m_\varphi\Big),
\end{align*}
}
where $m_\varphi=\frac{\rho^2}{f}-f\omega^2$.

Komar's angular momentum is defined as
{\small
\begin{align}\label{eq:MomentoAngular_Komar}
    J(x)&= \frac{1}{16\pi}\int_{S} *d(({}_\varphi K)^\flat)=\frac{1}{16\pi}\int_0^{2\pi} d\varphi \int_{-1}^{1} dy \Big[*d(({}_\varphi K)^\flat)\Big]_{y\varphi} \notag \\
    &=\frac{\lambda _0 L^2}{48 f_0\, x^3 \left(x^2+1\right)}\bigg[\lambda _0^2 \bigg(3 \left(x^2-1\right) \left(x^2+1\right)^2 \cot ^{-1}(x) \notag \\
    & -x \left(3 x^4+2 x^2+3\right)\bigg)+3 \left(x^2+1\right) \Big(\tau _0^2 \left(x^2+1\right) \ldots  \notag \\
    & \ldots \left(\left(x^2-1\right) \cot ^{-1}(x)-x\right)-8 x^3\Big)\bigg]
\end{align}
}
Therefore
\begin{equation}\label{eq:MomentoAngular_KomarInfinito}
    J_{\infty}=\lim_{x\to\infty}J(x)=-\frac{\lambda_0 \, L_{\pm}^2 }{2f_0}=-\frac{\lambda_0 \,L_{\pm}}{\tau_0 \,f_0}\, N_{\infty}.
\end{equation}

The electric charge $Q$ is defined as the flux of the displacement field, i.e., the surface integral of the 2-form $e^{-2\alpha_0\phi}\,\star F$ (the dual of $F$ with the dilatonic factor in EMD) \cite{Gibbons:1987ps,Garfinkle:1990qj,Ibadov:2020ajr}, while the magnetic charge $H$ is defined as the flux of $F$ through a closed 2-surface, i.e., the surface integral of $F$ (with Bianchi identity $dF=0$), using the same normalization and sign conventions as in standard EMD work and in comparative studies of wormholes with NUT \cite{Gibbons:1987ps,Garfinkle:1990qj,Clement:2015aka,Clement:2022pjr}.

From $A=A_t\,d t + L_{\pm } \, A_\varphi\,d\varphi$ we get
\begin{align*}
    F&=d A
=(\partial_x A_t)\,d x\wedge d t +(\partial_y A_t)\,d y\wedge d t \\
&+L_{\pm } \,(\partial_x A_\varphi)\,d x\wedge d\varphi +L_{\pm } \,(\partial_y A_\varphi)\,d y\wedge d\varphi.
\end{align*}
Thus, since $F_{tx}=-\partial_x A_t$ and $F_{\varphi x}=-L_{\pm}\,\partial_x A_\varphi$, and using \eqref{eq:HodgeHMaestra} to evaluate $(*F)_{y\varphi}$, we can derive the electric flux charge 
{\small
\begin{align}\label{eq:Q_Komar}
&Q(x)=\frac{1}{4\pi}\int_{S_x} e^{-2\alpha\phi}\,*F
=\frac{1}{4\pi}\int_0^{2\pi}\!\!d\varphi\int_{-1}^{1}\!\!d y\;
e^{-2\alpha\phi}\,(*F)_{y\varphi}\nonumber\\
&=\frac{L_{\pm}\, (x^2\pm1) }{2}\int_{-1}^{1}d y\;
e^{-2\alpha\phi}
\left[
\left(\frac{1}{f}-\frac{f\omega^2}{\rho^2}\right)\partial_x A_t
-\frac{f\omega}{\rho^2}\partial_x A_\varphi
\right]\notag \\
&=\frac{L_{\pm}\, \tau_0 \, \cosh \left(\frac{\lambda _0}{x^2+1}\right)}{2 \sqrt{\sigma_0 \, f_0} \, \kappa_0} e^{\frac{\tau_0 x}{x^2+1}}
\end{align}
}
Then, the electric flux charge at infinity is
\begin{equation}\label{eq:Q_KomarInfinito}
    Q_{\infty}=\lim_{x\to\infty}Q(x)=\frac{\tau_0 \, L_{\pm} }{2 \kappa_0 \sqrt{\sigma_0f_0}}.
\end{equation}

Finally, the magnetic flux charge is defines as
\begin{align}\label{eq:H_Komar}
H(S)&=\frac{1}{4\pi}\int_{S_x}F=\frac{L_{\pm}}{4\pi}\int_0^{2\pi}\!\!d\varphi\int_{-1}^{1}\!\!d y\;\partial_y A_\varphi \notag \\
&=\frac{L_{\pm}}{2}A_{\varphi}\bigg|_{y=-1}^{y=1}.
\end{align}
Then, the magnetic flux charge at infinity is
\begin{equation}\label{eq:H_KomarInfinito}
    H_{\infty}=\lim_{x\to\infty}H(x)=\frac{\tau_0 \, L_{\pm} }{2 \kappa_0 \sqrt{\sigma_0f_0}}=Q_{\infty}.
\end{equation}
\section{Distinction between compact objects}

\subsection{Event horizon}
To determine the event horizons, we define the Killing vector $K_t=\partial_t + \Omega\,\partial_\psi$, where $\Omega$ remains constant on the horizon $\mathscr{H}$. We then evaluate its norm, $g(K_t,K_t)=g_{tt}+2\Omega\, g_{t\psi}+\Omega^2 g_{\psi\psi}=0$, and solving this equation yields:
\begin{equation*}
    \Omega_{\pm }=-\frac{g_{t\psi}}{g_{\psi \psi}}\pm \frac{\sqrt{(g_{t\psi})^2-g_{\psi \psi}g_{tt}}}{g_{\psi \psi}}
\end{equation*}
By imposing $(g_{t\psi})^2 - g_{\psi\psi} g_{tt} = \rho =0$, then $L_{\pm}\sqrt{(x^2 \pm 1)(1-y^2)} = 0$ and setting $f=1$, we obtain the unique, non-degenerate value $\Omega = \Omega_- = \Omega_+=-g_{t\psi}/g_{\psi\psi} = \left(\omega\big|_{\mathscr{H}}\right)^{-1}$.

In the SU-E (Upper sign) case, \textit{no event horizon is present}, since the condition $\rho=0$ implies $y=\pm 1$, and plugging this into \eqref{Omega Lambda6} yields $\omega|_{\mathscr{H}}=\omega(x, y=\pm 1)=0$. 

On the other hand, in the S-E (Lower sign) configuration, \textit{the event horizon is located at:} 
\begin{equation}\label{Horizonte Eventos at}
    x=\pm 1 \Rightarrow r=l_1\pm L_{-},
\end{equation}
with $\omega|_{\mathscr{H}}=\omega(x=\pm L_{\pm},y)=L_{-}\lambda_0:\text{constant}$, the \textbf{sub-extreme configuration indeed corresponds to a Black Hole.}
The surface gravity at the event horizon is obtained from
\begin{equation}\label{Gravedad Superficial}
    \varsigma=\lim_{p\rightarrow\mathscr{H}}\sqrt{\gamma^{ij}(\partial_i\Xi)(\partial_j\Xi)}
=\frac{e^{-k_0(\lambda_0^2-\tau_0^2)/4}}{L_{-}\lambda_0},
\end{equation}
where $\Xi=\sqrt{g(K_t,K_t)}$, $\gamma^{ij}$ is the the metric restricted to the ($x,y$)-subspace and $\{i,j\}=x,y$.
\subsection{Wormhole throat}
The throat is defined as the minimum of the areal function given by
\begin{align}\label{Funcion Areal}
A_{\text{real}} &= \int_{0}^{2\pi } \int_{1}^{-1} \sqrt{g_{yy} g_{\varphi\varphi}}\, dy d\varphi \notag \\ 
&=2\pi (L_\pm)^2 \int_{1}^{-1} \frac{e^{k}}{f} \sqrt{\frac{x^2 \pm y^2}{x^2\pm 1}}\, dy.
\end{align}
In other words, the throat is located at $x = x_G$, where $\left.\frac{d A_{\text{real}}}{d x}\right|_{x_G} = 0$ and $\left.\frac{d^2 A_{\text{real}}}{d x^2}\right|_{x_G} > 0$.
To determine the location of the throat, we employ numerical methods to obtain the value $x = x_G$.

The numerical analysis gives important insights. 

First, wormholes do not form for $\lambda_0,\tau_0 \geq 1$. In contrast, when $\lambda_0 \leq 10^{-3}$, $\tau_0 \leq 10^{-4}$, $L_{+} = 2$, and $k_0 = 3/4$ (dilatonic field in superstring theory), they are stable and exhibit a throat in the SU-E case (upper sign). However, in the S-E case (lower sign), the throat is always confined to the region $|x_G|<1$. Thus, if we interpret the sub-extreme configuration as a wormhole, its throat is hidden behind the event horizon, and the wormhole geometry only appears in the super-extreme regime. In other words,the SU-E configuration corresponds to a wormhole (WH), while the S-E configuration corresponds to a black hole (BH).

Second, the throat (SU-E case) lies at $x = 0$ for $y \in \Big[-1, -90(\lambda_0+\tau_0)\Big) \cup \Big(90(\lambda_0+\tau_0), 1\Big]$ approximately. In contrast, for $y \in \Big[ -90(\lambda_0+\tau_0),90(\lambda_0+\tau_0)\Big]$, the throat is shifted to a position $x_G \neq 0$, where $x_G$ depends on the angle $y_0$. An example, when $y_0=0 \Rightarrow x_G \approx 0.0293$.

\begin{figure}[b]
    \centering
    \begin{minipage}{0.45\textwidth}
    \centering
        \includegraphics[width=\textwidth]{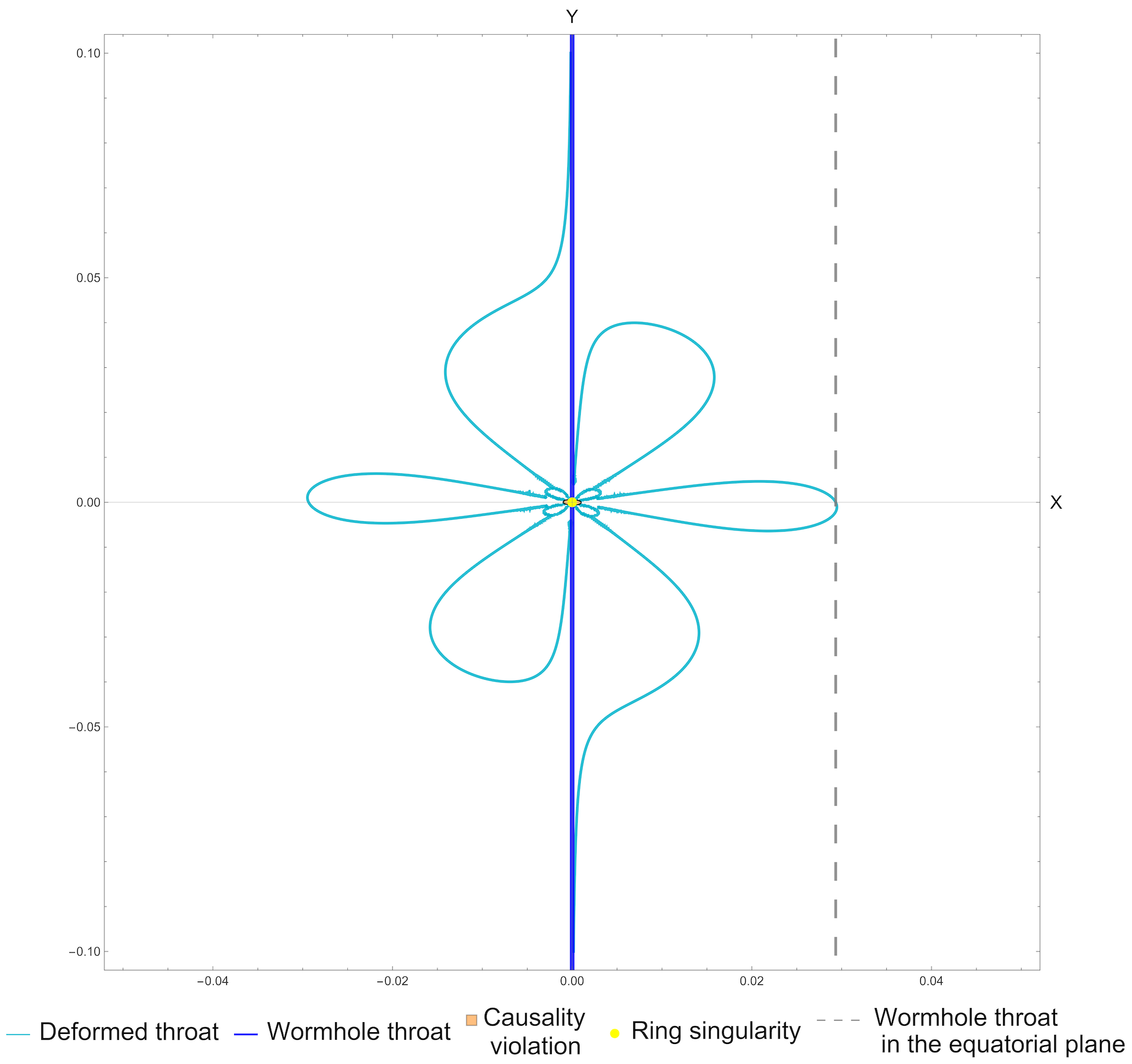}
        \subcaption{Shape of the deformed WH throat.}
        \label{fig:Garganta1}
    \end{minipage}
    \hfill
    \begin{minipage}{0.45\textwidth}
    \centering
    \includegraphics[width=\textwidth]{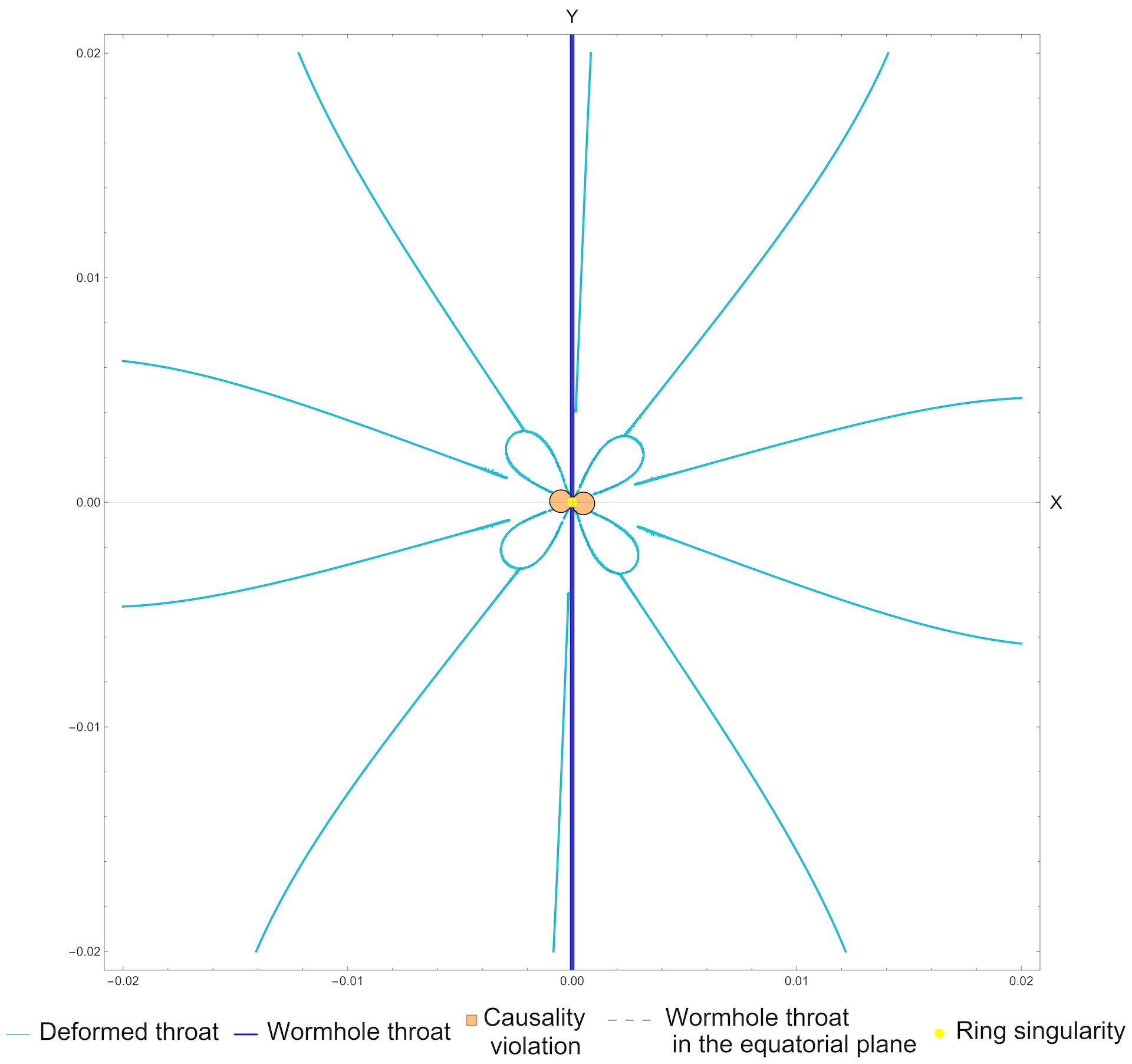}
        \subcaption{Zoom-in on the wormhole in the vicinity of causality anomalies.}
        \label{fig:Garganta2}
    \end{minipage}
    \caption{The parameters used were $\lambda_0=10^{-3}$, $\tau_0=10^{-4}$, $k_0=3/4$, $L=2$. Over all the fuchsia line (deformed wormhole throat) the conditions $d A_{\text{real}}/dx|_{x_G}=0$ and $d^2 A_{\text{real}}/d^2x|_{x_G}>0$ is fulfill. }\label{fig:GargantaWormhole}
\end{figure}
In Figure \ref{fig:GargantaWormhole}, we present a comparison between the wormhole throat at $x_G = 0$, shown as a blue continuous line when $y \notin \Big[ -90(\lambda_0+\tau_0), 90(\lambda_0+\tau_0)\Big]$, and the deformed wormhole throat with $x_G \neq 0$, represented by a fuchsia continuous line for $y \in \Big[ -90(\lambda_0+\tau_0), 90(\lambda_0+\tau_0)\Big]$.On the other hand, we plot the orange region corresponding to the causal violations, i.e., the area where CTCs appear, and the ring singularity is indicated by the yellow point at $x = y = 0$. Finally, the dashed gray line represents the throat in the equatorial plane $y = 0$ with $x_G \approx 0.0293$. 

\subsection{Closed Time-like Curves}
To identify CTCs, note that within the coordinate range $\varphi \thicksim \varphi + n\pi$ there is a closed curve. Consequently, the coordinate $\varphi$ serves as a time function whenever $g(\partial_\varphi,\partial_\varphi)=g_{\varphi \varphi}=(\rho^2-f^2\omega^2)/f<0$.  
In the BH case (S–E), causality violations arise in the region $|x|<1$ if and only if $\lambda_0,\tau_0<1$. When $\lambda_0,\tau_0 \geq 1$, however, these violations shift to certain zones with $|x|>1$ (outside the event horizon).  
For the WH configuration (SE–E), causality violations occur in the domain specified by $|x_v| \approx \lambda_0 + \tau_0 \ll 1$ and $y\in [-(\lambda_0+\tau_0)/2,(\lambda_0+\tau_0)/2]$. Yet, if $\lambda_0 \leq 10^{-3}$, $\tau_0 \leq 10^{-4}$ and we fix $L_{+}=2,k_0=3/4$, then $|x_v|<|x_G|$. In other words, the throat encompasses all irregularities, assuming a role analogous to that of horizons in a black hole. Therefore, this WH satisfies the WCCC, as explained in \cite{axioms14110831}.
A graphical illustration is provided in Figure \ref{fig:GargantaWormhole}.
\section{Null Energy Condition}
\label{Null energy condition}

The Null Energy Condition (NEC) is determinate by the following inequality
\begin{equation*}
    T_{\mu \nu} l^{\mu} l^{\nu} \geq 0, 
\end{equation*}
where $l^{\mu}$ is the null vector, i.e., satisfies $l^{\mu}l_{\mu}=0$, and $\tensor{T}{_{\mu \nu}}$ is the energy-momentum tensor.
To study the NEC, it is necessary to define the comoving reference system $\mathbb{O}$, which physically corresponds to a frame fixed to the compact object. For this purpose, we will use a diagonal tetrad. The matrix transformation of the coordinate basis of the tangent space to $\mathbb{O}$ is given by
{\small
\begin{equation}\label{MatrizDeTransformacionComovil}
    \mathbb{M}_{comoving}^{T}=\begin{bmatrix}
        \begin{array}{cccc}
             1/\sqrt{-\tensor{g}{_{t t } }} & 0 & 0 & 0 \\
             0 & 1/\sqrt{\tensor{g}{_{x x } }} & 0 & 0 \\
             0 & 0 & 1/\sqrt{\tensor{g}{_{y y } }} & 0 \\
             \frac{-\tensor{g}{_{\varphi t}}/\tensor{g}{_{t t}} }{{Det_{\varphi t}} }& 0 & 0 & \frac{1}{Det_{\varphi t}}
        \end{array}
    \end{bmatrix}.
\end{equation} }
%
where $Det_{\varphi t}=\sqrt{\tensor{g}{_{\varphi \varphi}}-(\tensor{g}{_{\varphi t}})^2/\tensor{g}{_{t t}}}$.
Thus, the relationship between the basis vectors in $\mathbb{O}$ and the coordinate basis is
\begin{equation}\label{RelacionDeBasesComovilMetricaTesis}
    \begin{bmatrix}
    \begin{array}{c}
     \partial_0 \\
     \partial_1 \\
     \partial_2 \\
     \partial_3
    \end{array}
    \end{bmatrix}^{T}
    =\mathbb{M}_{comoving}^{T}
    \begin{bmatrix}
    \begin{array}{c}
     \partial_t \\
     \partial_x \\
     \partial_y \\
     \partial_{\varphi}
    \end{array}
    \end{bmatrix}^{T}.
\end{equation}
We use the indices $a,b=0,\dotsi,3$ for $\mathbb{O}$ and $\alpha,\beta,...$ for our coordinate system, so the Ricci tensor transforms as
\begin{equation*}
    \tensor{R}{_{ab}}=\mathbb{M}_{comoving} \cdot \{R_{\mu \nu} \}\cdot (\mathbb{M}_{comoving})^{T}.
\end{equation*}
We choose the null vector as $l_{a}=\partial_{0}+\partial_{1}$ and rewrite the NEC in the $\mathbb{O}$-frame, obtaining the following form
{\setlength{\abovedisplayskip}{0pt}
 \setlength{\abovedisplayshortskip}{0pt}
\begin{multline}\label{Eq: De la condicion de energía}
    T_{a b} l^{a} l^{b}= T_{00}+ T_{11}=\varrho-P \\ =\frac{c^4}{8 \pi G}\bigg( R_{00}+ R_{11} \bigg) \geq 0,
\end{multline}
}
where $\varrho$ is the density and $P$ the pressure.

For the second class of solutions (\ref{SegundaClaseSoluciones}), we find that the density satisfies the following condition
\begin{multline}\label{Densidad 2da familia de soluciones}
    \frac{8\pi G}{c^4}\varrho =R_{00}= \frac{e^{-2k}}{2L_\pm ^4 (x^2\pm 1) (1-y^2) (x^2 \pm  y^2)} \\ \bigg( (1-y^2) \omega_{,y}^2 + (x^2\pm 1) \omega_{,x}^2 \bigg) \geq 0,
\end{multline}
Observe that there are only quadratic terms in (\ref{Densidad 2da familia de soluciones}), for this second class of solutions, one always has $\varrho \geq 0$ for the WH, and for the BH this holds if and only if $x^2 > 1$.

On the other hand, the expression for the pressure is
{\setlength{\abovedisplayskip}{0pt}
 \setlength{\abovedisplayshortskip}{0pt}
\begin{multline}\label{Presion 2da familia de soluciones}
        -\frac{8\pi G}{c^4} P= \tensor{R}{_{11}}= \frac{e^{-2k}}{L_\pm ^2 (x^2 \pm y^2)} \bigg( -(1-y^2) k_{,yy} \\- (x^2\pm 1) k_{,xx} +2y k_{,y} + \frac{\omega_{,x}^2}{2L_\pm ^2 (1-y^2)} \bigg) ,
\end{multline}
}
for constant scalar field and rotation, $P=0$. The explicit expression of the NEC \eqref{rho - varrho Lambdac} is presented in Appendix \ref{NEC in the diagonal tetrad}.

In the following, we will analyse the NEC (\ref{rho - varrho Lambdac}). Initially, it is imperative to obtain the expression that represents the extreme values of $x$ and $y$, similarly, the intermediate regions corresponding to $x \approx y$. It is worth noting that $y\in [-1,1] \quad$ implies $\quad y^{2n} \in [0,1]$ appears in one case, $x \in (-\infty, \infty) \quad $ implies $ \quad x^{2n} \in [0,\infty)$, then
{\small
{\setlength{\abovedisplayskip}{5pt}
 \setlength{\abovedisplayshortskip}{0pt}
\begin{subequations}\label{ComportamientoAsimptoticoRhoMenosP}
    \begin{equation}\label{RhoMenosPInfinito}
        \text{For} \quad x \gg 1 \,\,\, \text{implies}\,\,\, (\varrho-P)\approx \frac{(4k_0+1) \tau_0^2}{2L_\pm ^2x^4},
    \end{equation}
    \begin{equation}\label{RhoMenosPXCero}
        \lim\limits_{x \rightarrow  0 } (\varrho-P)= \frac{e^{-2 k_c(0,y)}}{2 L_\pm ^2 y^6} \Big\{ \pm 2\lambda_0^2 (1-y^2) +\tau_0^2 (4k_0+1) \Big\},
    \end{equation}
    \begin{equation}\label{RhoMenosPy0}
        (\varrho-P)_{|y=0}= \frac{e^{-2 k_c(x,0)}}{2 L_\pm ^2 x^6} \Big\{ 2\lambda_0^2 +\tau_0^2 (4k_0 +1) (x^2\pm 1)  \Big\},
    \end{equation}
    \begin{equation}\label{RhoMenosPy1}
        (\varrho-P)_{|y=1}= (4k_0+1)\frac{(2\lambda_0 x+\tau_0 \{ x^2\mp 1\})^2}{2L_\pm ^2(x^2\pm 1)^4}.
    \end{equation}
    
\end{subequations}
}
}
We need to examine the cases individually; for the WH (Upper) scenario, we consider
{\setlength{\abovedisplayskip}{0pt}
 \setlength{\abovedisplayshortskip}{0pt}
\begin{multline}\label{RhoMenosPXY}
        \lim\limits_{x \rightarrow  y } (\varrho-P)= \frac{e^{-2 k_c(y,y)}}{16 L_+^2 y^6} \Big\{ 2\tau_0^2 (1-y^2) \\
        +\lambda_0^2 (4k_0+1) (1+y^2) \Big\},
    \end{multline}
}
and for the BH (Lower) scenario
{\setlength{\abovedisplayskip}{0pt}
 \setlength{\abovedisplayshortskip}{0pt}
\begin{multline}\label{RhoMenosPX1}
        \lim\limits_{x \rightarrow  1 } (\varrho-P)= \frac{e^{-2 k_c(1,y)}}{L_{-}^2 (1-y^2)^4} \Big( \lambda_0 (1+y^2)+2\tau_0 y\Big)^2.
    \end{multline}
}
From Eqs. \eqref{RhoMenosPInfinito} and \eqref{RhoMenosPy1}, it follows that, in the asymptotic regime ($x \gg 1$) and along the polar axis, the term $(4k_0+1)$ provides the leading contribution in determining whether the NEC is fulfilled, for both WH and BH configurations.
The reason is that, in the limit $x \rightarrow \infty$, $(\varrho - P)$ tends to $0^{\pm}$, and the result depends on the sign of $(4k_0 + 1)$, i.e., whether $(4k_0 + 1) > 0$ or $(4k_0 + 1) < 0$. In particular, whether the limit goes to zero from the positive side, $0^{+}$, or from the negative side, $0^{-}$, determines if the NEC is satisfied or violated.

In the BH case, we observe that the expressions \eqref{RhoMenosPy0} and \eqref{RhoMenosPX1} are non-negative provided that we restrict the analysis to the domain $x \geq 1$ and impose the condition $(4k_0 + 1) \geq 0$.This domain includes both the event horizon itself and the outer region, in a spacetime setting permeated by a dilatonic scalar field.

Conversely, since an exponential function is strictly positive and we have $\lambda_0, \tau_0 \in \mathbb{R}$ with $y \in [-1,1]$, it follows that $y^{2n} \in [0,1]$, implying $1 - y^2 > 0$. Therefore, for the WH in the intermediate region where $x \approx y$ and on the equatorial plane, equations (\ref{RhoMenosPXY}) and \eqref{RhoMenosPy0} indicate that the NEC are fulfilled whenever $(4k_0 + 1) > 0$.

Therefore, we find that for WHs, and for BHs only when $x \geq 1$, the following holds

{\setlength{\abovedisplayskip}{-2pt}
 \setlength{\abovedisplayshortskip}{-2pt}
\begin{align*}
    &\textbf{If} \quad 4k_0+1 >0 \quad \,\,\, \text{implies}\,\,\, \quad (\varrho-P)>0,  \\
    &\textbf{If} \quad 4k_0+1 <0  \quad \,\,\, \text{implies}\,\,\, \quad (\varrho-P)<0.
\end{align*}
}

In summary, we have obtained the known results for WHs as well as for BHs (both at the event horizon and in the exterior region):

{\setlength{\abovedisplayskip}{-2pt}
 \setlength{\abovedisplayshortskip}{-2pt}
\begin{align*}
    &\textbf{Dilatonic Field} \quad \,\,\, \text{implies}\,\,\, \quad (\varrho-P)>0,  \\
    &\textbf{Phantom Field}  \quad \,\,\, \text{implies}\,\,\, \quad (\varrho-P)<0.
\end{align*}
}

\section{Petrov Classification}

To determine the Petrov type of the spacetime, we constructed a Newman-Penrose
null tetrad ($\ell^{\mu} = \frac{1}{\sqrt{2}}\left(\partial_{0}^{\mu} + \partial_{1}^{\mu}\right)$, $n^{\mu} = \frac{1}{\sqrt{2}}\left(\partial_{0}^{\mu} - \partial_{1}^{\mu}\right)$, $ m^{\mu} = \frac{1}{\sqrt{2}}\left(\partial_{2}^{\mu} + i\,\partial_{3}^{\mu}\right)$ and $\bar{m}^{\mu} = \frac{1}{\sqrt{2}}\left(\partial_{2}^{\mu} - i\,\partial_{3}^{\mu}\right)$) adapted to the orthonormal frame introduced in Sec.~\ref{Null energy condition}. Using the standard contractions of the Weyl tensor with this tetrad, we computed the five Newman--Penrose Weyl scalars $\Psi_0,\dots,\Psi_4$:
{\small
\begin{align*}
    &\Psi_0 = - C_{abcd} \, \ell^a m^b \ell^c m^d, \quad
    \Psi_1 = - C_{abcd} \, \ell^a n^b \ell^c m^d, \\
    &\Psi_2 = - C_{abcd} \, \ell^a m^b \bar{m}^c n^d, \quad
    \Psi_3 = - C_{abcd} \, \ell^a n^b \bar{m}^c n^d, \\
    &\Psi_4 = - C_{abcd} \, \bar{m}^a n^b \bar{m}^c n^d,
\end{align*}
}
where $C_{abcd}$ is the Weyl tensor.
For our solution, all five scalars are non-vanishing functions of $(x,y)$, so the spacetime is not of type II, III, N, or D.

We then formed the usual algebraic invariants
{\small
\begin{equation}
    I \;=\; \Psi_0 \Psi_4 - 4 \Psi_1 \Psi_3 + 3 \Psi_2^2 \, , 
    \quad
    \mathcal{J} \;=\; \det
    \begin{pmatrix}
    \Psi_4 & \Psi_3 & \Psi_2 \\
    \Psi_3 & \Psi_2 & \Psi_1 \\
    \Psi_2 & \Psi_1 & \Psi_0
    \end{pmatrix},
\end{equation}
}
which are both nonzero for generic values of the parameters and 
coordinates. Moreover, we find that $I^3 \neq 27 \mathcal{J}^2$, so that the Weyl tensor is algebraically general (Petrov type I), according to the standard invariant classification \cite{Newman:1961qr,Pirani:1956wr,Cita:ExactSolutions,Zakhary:1997xas}.
The explicit expressions of the scalars $\Psi_i$ and of the invariants $I$ and
$\mathcal{J}$ are very lengthy and will not be displayed here,but in the next subsection, we will present the asymptotic behavior of the NPW scalars. 


\subsection{Asymptotical analysis}
Another key step is to compute the Maxwell scalars ($\Phi_{0} = F_{ab}\,\ell^{a}m^{b}$, $\Phi_{1} = \tfrac{1}{2} F_{ab}(\ell^{a}n^{b} + \bar{m}^{a}m^{b})$, $\Phi_{2} = F_{ab}\,\bar{m}^{a}n^{b}$) using the Newman–Penrose null tetrad, yielding (The upper symbol represents the WH, and the lower one represents the BH):
{\small
\begin{align}\label{Escalares de MAxwell}
    &\Phi_0=-\Phi_2 \notag \\ &=\mp \frac{(i-1)\sqrt{1-y^2}}{4L_{\pm } (x^2\pm y^2)^{5/2}}\Big(\lambda_0 (\mp x^2+y^2) +2\tau_0 xy \Big) e^{-k-\lambda_c}, \\
    &\Phi_1=\mp \frac{(i-1)\sqrt{x^2\pm 1}}{4L_{\pm} (x^2\pm y^2)^{5/2}}\Big(\tau_0 (x^2\mp y^2) +2\lambda_0 xy \Big) e^{-k-\lambda_c}.
\end{align}
}
If we consider the asymptotic regime $x\rightarrow\pm\infty$ (with $L_\pm x \simeq r$), the Newman–Penrose Weyl scalars behave as $\Psi_0=\Psi_4 \simeq i \tau_0 (1-y^2) (k_0 \tau_0^2 \pm 3)/(4 L_\pm ^2 x^5)$, $-\Psi_1=\Psi_3 \simeq 3 i \lambda_0 \sqrt{1-y^2}/(4 L_\pm ^2 x^4)$, and $\Psi_2 \simeq -i \tau_0 /(2L_\pm ^2x^3)$. Thus, the dominant Coulomb–like contribution to the curvature is determined by the parameter $\tau_0$, while the next–to–leading multipolar structure is controlled by the constant $\lambda_0$. Similarly, the Maxwell scalars obey $\Phi_1 \simeq (i-1)\tau_0/(4L_\pm x^2)$ and $\Phi_0=-\Phi_2 \simeq (i-1)\lambda_0\sqrt{1-y^2}/(4L_\pm x^3)$, indicating that the net dyonic Coulomb field is again governed by $\tau_0$, whereas $\lambda_0$ encodes the higher–order electromagnetic multipoles. This ordering in the fall–off rates, with the Coulombic pieces $\Psi_2$ and $\Phi_1$ decaying more slowly than $\Psi_{0,1,3,4}$ and $\Phi_{0,2}$, is typical of a nonradiative, Coulomb–dominated configuration in the Newman–Penrose framework \cite{Newman:1961qr,Newman:1962cia,Penrose:1962ij,GomezLopez:2017kcw}.

In the NUT/ALF branch ($\tau_0\neq 0$), the Coulombic components $\Psi_2$ and $\Phi_1$ are nonvanishing and are completely fixed by the value of $\tau_0$. In contrast, in the asymptotically flat branch ($\tau_0 = 0$), these Coulomb-type pieces disappear, and only the higher–multipole tails $\Psi_1,\Psi_3$ and $\Phi_0,\Phi_2$ remain. Consequently, the spacetime tends toward a truly asymptotically flat regime, free of NUT charge and containing purely multipolar fields. This yields an invariant, peeling–theoretic characterization of the distinct asymptotic behavior in the ALF ($\tau_0\neq 0$) versus AF ($\tau_0=0$) sectors of the solution.
\section{Wormhole Geometry}
To analyze WH geometry, the use of embedding diagrams is essential. Initially, the WH is embedded in a cylindrical plane using two different methods. These methods involve selecting a hypersurface with parameter $t=t_0$ and another variable kept constant, depending on the desired shape visualization. Setting $y=y_0$ yields a diagram illustrating the throat connecting two regions. Conversely, using $x=x_0$ yields the throat shape. Examples of this procedure, along with detailed explanations, can be found in \cite{Lobo:2017cay} and \cite{DelAguila:2023twe}.

In our case $f=1$, then we take $t=t_0$ and $y=y_0$ in (\ref{ds sp}), we arrive to

{\setlength{\abovedisplayskip}{-2pt}
 \setlength{\abovedisplayshortskip}{-2pt}
\begin{multline}\label{ds Hipersuperficie ycte}
    ds^2 =  L_+^2\frac{(x^2+y_0^2) e^{2k(x,y_0)}}{x^2+1}  dx^2 \\+\left[ L_+^2 (x^2+1)(1-y_0^2) -\omega(x,y_0)^2 \right] d\varphi^2.
\end{multline}
}

Fixing $t=t_0$ and $x=x_0$ in (\ref{ds sp}), we acquired

{\setlength{\abovedisplayskip}{-2pt}
 \setlength{\abovedisplayshortskip}{-2pt}
\begin{multline}\label{ds Hipersuperficie xcte}
    ds^2 =  L_{+}^2\frac{(x_0^2+y^2) e^{2k(x_0,y)}}{1-y^2}  dy^2 \\+\left[ L_+^2 (x_0^2+1)(1-y^2) -\omega(x_0,y)^2 \right] d\varphi^2.
\end{multline}
}

\begin{figure}[b]
    \centering
    \begin{minipage}{0.48\textwidth}
    \centering
        \includegraphics[width=\textwidth]{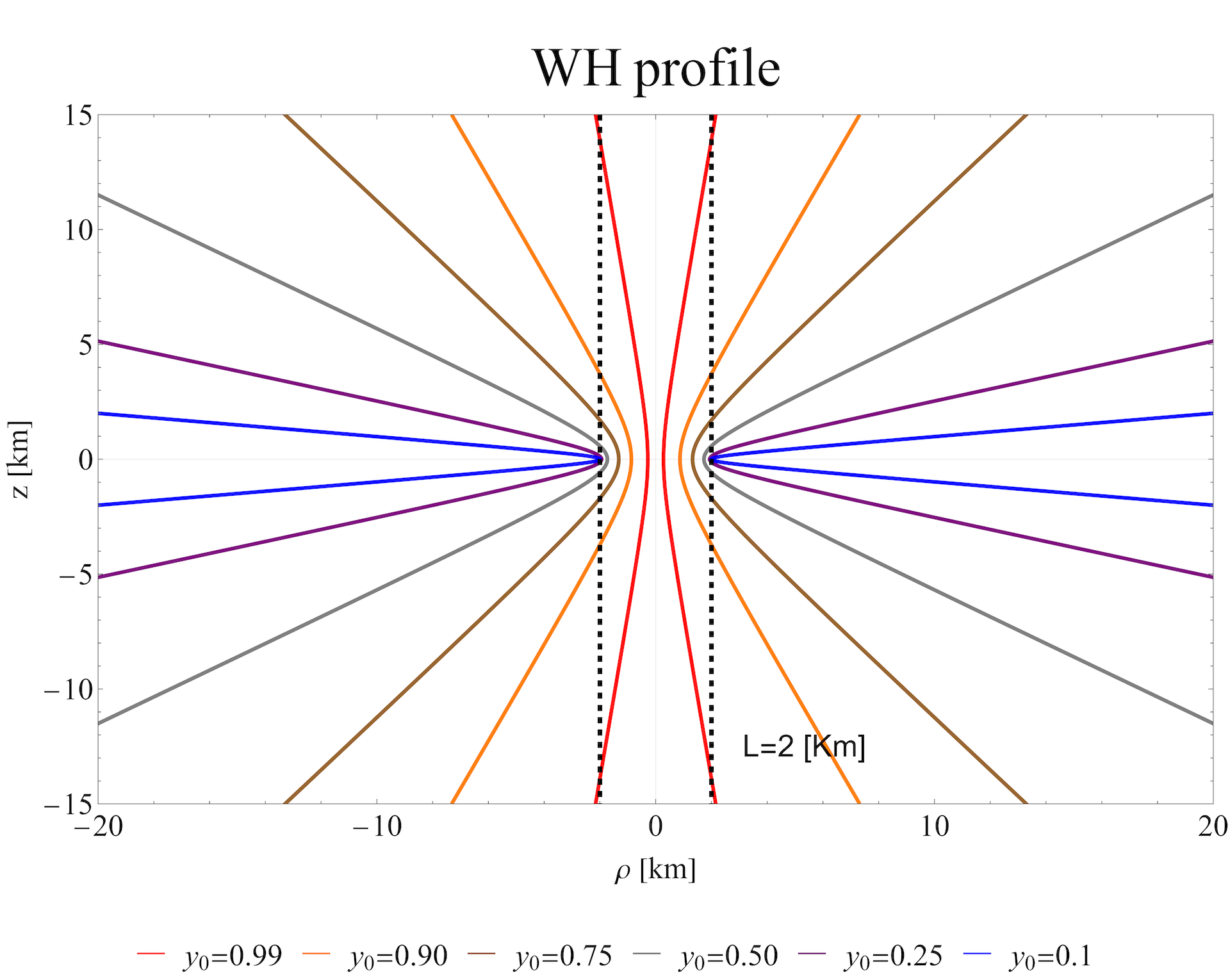}
        \subcaption{Profile of the WH.}
        \label{fig:PerfilWH}
    \end{minipage}
    \hfill
    \begin{minipage}{0.48\textwidth}
    \centering
    \includegraphics[width=\textwidth]{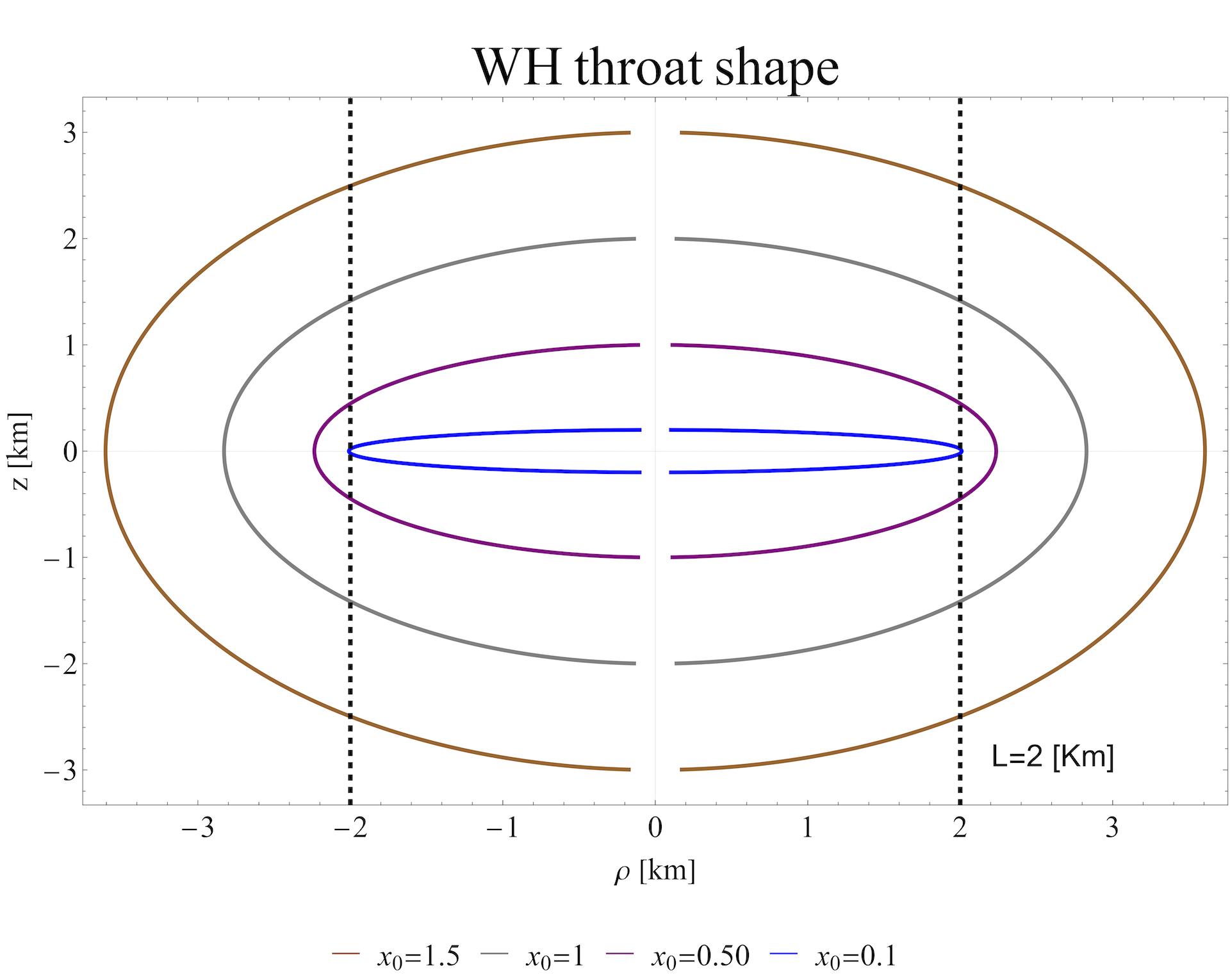}
        \subcaption{Shape of the WH throat.}
        \label{fig:FormaGarganta}
    \end{minipage}
    \caption{Both graphs use the same parameters: $\lambda_0 = 10^{-3}$, $\tau_0 = 10^{-4}$, $k_0 = 3/4$, and a WH with the size of the Sun. The initial condition used for the numerical solution is $z(x=0)$. Furthermore, all units in the graphs are expressed in kilometers.}\label{fig:ShapeProfileWH}
\end{figure}

We will use a flat cylindrical space representation $(x,y)$, and consider a parameterization with respect to $x^i$ at constant $x^j \neq x^i$, where if $x^i=x$ implies $x^j=y_0$ or $x^i=y$ implies $x^j=x_0$
{\small
\begin{align}
    ds^2&=d\overline{\rho}^2+dz^2+\overline{\rho}^2d\varphi^2 \nonumber \\
    &=\left\{ \left( \frac{d\overline{\rho}}{dx^i} \right)^2 +\left( \frac{dz}{dx^i} \right)^2 \right\}(dx^i)^2+\overline{\rho}(x^i,(x_0)^j)^2 d\varphi^2, \label{CilindricasParametrizadasHipersup}
\end{align}
}
where $\overline{\rho}$ is the corresponding radial variable in the coordinates of the cylindrical plane.
It is possible to embed the hypersurface (\ref{ds Hipersuperficie ycte}) or (\ref{ds Hipersuperficie xcte}) in (\ref{CilindricasParametrizadasHipersup}), that is, to equate the corresponding metric elements to derive differential equations, which will be solved numerically and subsequently visualized in a figure \ref{fig:ShapeProfileWH}. 
The 2 differential equation systems that arise from the embedding, are
\begin{subequations}\label{Ec de la GoemtriaHipersup ycte}
    \begin{align}
        &\overline{\rho}(x,y_0)^2= L_+^2 (x^2+1)(1-y_0^2) -\omega(x,y_0)^2  , \label{rho(x) ycte} \\
        &\left( \frac{d\rho}{dx} \right)^2 +\left( \frac{dz}{dx} \right)^2= L_+^2\frac{(x^2+y_0^2)}{x^2+1} e^{2k(x,y_0)}. \label{EcDif rhoZ(x) ycte}
    \end{align}
\end{subequations}
\begin{subequations}\label{Ec de la GoemtriaHipersup xcte}
    \begin{align}
        &\overline{\rho}(x_0,y)^2= L_+^2 (x_0^2+1)(1-y^2) -\omega(x_0,y)^2  , \label{rho(x) xcte} \\
        &\left( \frac{d\rho}{dy} \right)^2 +\left( \frac{dz}{dy} \right)^2= L_+^2\frac{(x_0^2+y^2)}{1-y^2} e^{2k(x_0,y)}. \label{EcDif rhoZ(x) xcte}
    \end{align}
\end{subequations}

By solving numerically and obtaining $\overline{\rho} (x),\overline{\rho} (y)$ and $z(x),z(y)$, we can graph $z(\overline{\rho})$ with the initial condition. 

\begin{figure}
    \centering
    \includegraphics[width=1\linewidth]{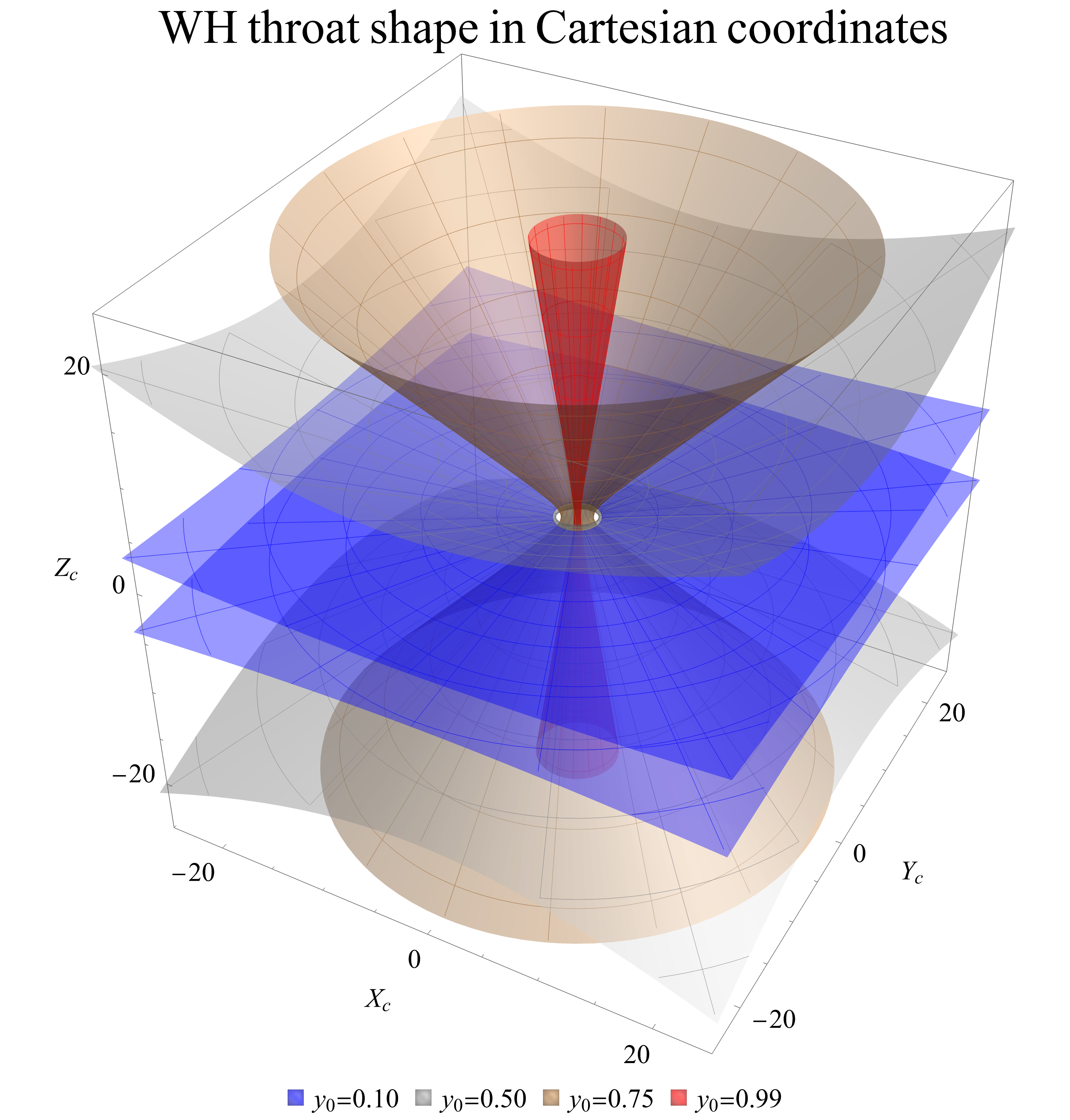}
    \caption{Hypersurface characterizing the WH profile. It is constructed as a surface of revolution derived from the curves illustrated in [\ref{fig:PerfilWH}].}
    \label{fig:GeometriaHW3D}
\end{figure}

Figure \ref{fig:PerfilWH} illustrates several curves, each graphed in a different color, that represent the embedding hypersurface for different values according to $y_0=\cos{\theta_0}$. These correspond to the numerical solution presented in (\ref{Ec de la GoemtriaHipersup ycte}). The black dashed line indicates the size of the WH as specified by $L=2$ Km, holding the condition $L>l_1$. It can be observed that, in the vicinity of the equatorial plane, the geometry assumes a flat configuration, while at the poles it resembles a cylindrical shape.
To understand the three-dimensional shape, one can examine Figure \ref{fig:GeometriaHW3D}, which illustrates the surface representation of each curve shown in Figure \ref{fig:PerfilWH}, now represented in three dimensions.

As illustrated in Figure \ref{fig:FormaGarganta}, the shape of the throat is represented by curves of different colors, each corresponding to discrete values of $x_0$, a value that approaches $L$ under the condition $x=0$. Similarly, the figure demonstrates that the configuration becomes flat as we approach the point $x=0$; this implies that as the system approaches the ring singularity, the WH geometry tends to close. The figure illustrates a discontinuity in the curves, denoted by the value $z=0$, this is because the system of differential equations (\ref{Ec de la GoemtriaHipersup xcte}) is not defined at the point $y=1$.

The last plot presented in Figure \ref{fig:GeometriaHW3D} shows the three-dimensional graphs corresponding to the surfaces of revolution generated by the curves in \ref{fig:PerfilWH}, evaluated for various values of $y_0$. It should be noted that not all curves are included to avoid overloading the illustration.

The radial coordinates determining the size of the WH in the embedding diagram are given by (\ref{rho(x) ycte}). However, in this case, the throat size depends on the angular position, given by $y$.
%

\section{WH Tidal Forces}

Following the methodology proposed in \cite{Morris:1988cz} (see also \cite{DelAguila:2015isj}, and \cite{Bixano:2025jwm}), the most appropriate reference frame for analyzing tidal forces is that of an astronaut $\hat{\mathbb{O}}$ attempting to cross the WH.
For simplicity, in the astronaut's reference frame, we will consider his 4-velocity to be purely temporal $\vartheta^{\hat{\mu}}=(1,0,0,0)$, which guarantees the orthogonality of the 4-acceleration and the 4-velocity $W^{\hat{\mu}} \vartheta_{\hat{\mu}}=0 $, which implies that $W^{\hat{\mu}}=(0,\hat{a},0,0)$. 
Therefore, by using the geodesic deviation equation
\begin{equation}\label{Tidal Forces}
    \Delta a^{\hat{\mu}}=-c^2 \tensor{R}{^{\hat{\mu}}}{_{\hat{\alpha} \hspace{0.2em} \hat{\beta} \hspace{0.2em} \hat{\sigma} }} \tensor{\vartheta}{^{\hat{\alpha}}} \tensor{\xi}{^{\hat{\beta}}} \tensor{\vartheta}{^{\hat{\sigma}}},
\end{equation}
It is possible to determine the danger zones where the astronaut can be destroyed by tidal forces. Here $\xi^{\hat{\mu}}$ is the 4-distance between the astronaut's head and feet, and $\tensor{R}{^{\hat{\mu}}}{_{\hat{\alpha} \hspace{0.2em} \hat{\beta} \hspace{0.2em} \hat{\sigma} }}$ are the components of the Riemann tensor in the astronaut's reference frame.
The Riemann tensor in $\hat{\mathbb{O}}$ is derived using the linear transformation matrix used in special relativity, which facilitates the transformation\footnote{The transformation matrix (\ref{MatrizTransformaciónRelEsp}) only transforms the covariant components.} from $\mathbb{O}$ to $\hat{\mathbb{O}}$.
\begin{equation}\label{MatrizTransformaciónRelEsp}
    \mathbb{M}_{RE}=
    \begin{bmatrix}
        \begin{array}{cccc}
            \beta & -\gamma \beta & 0 & 0 \\
            -\gamma \beta & \beta & 0 & 0 \\
            0 & 0 & 1 & 0 \\
            0 & 0 & 0 & 1 
        \end{array}
    \end{bmatrix},
\end{equation}
where $\beta= \vartheta /c $, $\vartheta$ is the speed of $\hat{\mathbb{O}}$ with respect of $\mathbb{O}$, and $\gamma= 1/\sqrt{1-\beta^2}$.

The relations of the Riemann component tensors are
{\setlength{\abovedisplayskip}{5pt}
 \setlength{\abovedisplayshortskip}{0pt}
\begin{subequations}\label{Riemman en SR del astronauta}
\begin{equation}\label{Rrtrt en SR del astronauta}
    \tensor{R}{_{ \hat{1} \hspace{0.2em} \hat{0} \hspace{0.2em} \hat{1} \hspace{0.2em} \hat{0} }}  =\tensor{R}{_{ 1010 }},
\end{equation}
\begin{equation}\label{R2t2t en SR del astronauta}
    \tensor{R}{_{ \hat{2} \hspace{0.2em} \hat{0} \hspace{0.2em} \hat{2} \hspace{0.2em} \hat{0} }} = \gamma^2 \tensor{R}{_{ 2020 }} + \gamma^2 \beta^2 \tensor{R}{_{ 2121}},
\end{equation}
\begin{equation}\label{R3t3t en SR del astronauta}
    \tensor{R}{_{ \hat{3}\hspace{0.2em} \hat{0} \hspace{0.2em} \hat{3} \hspace{0.2em} \hat{0} }}=\gamma^2 \tensor{R}{_{ 3030 }} + \gamma^2 \beta^2 \tensor{R}{_{ 3131 }}.
\end{equation}
\end{subequations}

Let $|\xi| \approx 2 m$ be the astronaut's height, and we consider that the maximum gravity that the human body can withstand is that of the Earth $g_{Earth}$, from which the following inequality is derived
\begin{equation}\label{CondicionFuerzasDeMarea}
    |\tensor{R}{_{\hat{\mu} \hspace{0.2em} \hat{0} \hspace{0.2em} \hat{\mu} \hspace{0.2em} \hat{0} }}| \leq \frac{g_{Earth}}{2m * c^2}\approx (10^5\text{Km})^{-2}.
\end{equation}
If the astronaut travels at non-relativistic speeds ($\beta \approx 0$ this implies that $\gamma \approx 1$), then $\tensor{R}{_{\hat{\mu} \hspace{0.2em} \hat{0} \hspace{0.2em} \hat{\mu} \hspace{0.2em} \hat{0} }} \approx \tensor{R}{_{ \mu0\mu0 }}$. Otherwise, it is necessary to consider the terms $\tensor{R}{_{ \mu 1 \mu 1 }} $. The explicit expressions of the Riemann tensor components in the astronaut frame are provided in Appendix \ref{Riemman en astronautFrame}.

To graph tidal forces, we restrict our consideration to a spacecraft exhibiting non-relativistic velocities, thus employing only the equations (\ref{R2121-Lambdac}) - (\ref{R4141-Lambdac}) in (\ref{Riemman en SR del astronauta}). For this purpose, we use a sun-sized ($ l_1\sim 1.5 $km) WH given by:
\begin{equation}\label{Condicion L SunSize}
    L=2000 m > l_1=r_s^{Sun}/2 \approx 1.5 \times 10^3  \text{m}.
\end{equation}
In Figure \ref{fig:FuerzasdeMarea}, the radial ($R_{1010}$) and $\varphi$-angular ($R_{3030}$) tidal forces in spheroidal oblate coordinates are shown as colored surfaces. The vertical axis $z$ indicates the magnitude of the associated tidal force, expressed in units of kilometers$^{-2}$. In both panels, it is evident that the region in which the wormhole can be traversed with the greatest degree of safety is located in the vicinity of the polar axis, $y \approx \pm 1$. It should be emphasized that the functional form and qualitative behavior of the radial tidal force, $R_{1010}$, are analogous to those of the $\theta$-angular tidal force, $R_{2020}$. For this reason, the latter is not shown. 

The yellow continuous curve represents the tidal forces evaluated at the wormhole throat, whereas the other continuous curves, plotted in various colors, correspond to the tidal forces of the null geodesics obtained by following the same procedure described in the next section. The parameter values employed to compute the trajectories are $L = 2$, $\lambda_0 = 10^{-3}$, $\tau_0 = 10^{-4}$, and $k_0 = 3/4$. The initial conditions are specified as $x(0) = 1$, $l_z = 0$, $p_y(0) = 1$, and $E = 1$, while the initial values $-p_x(0)=\{ 1.23,1.29,1.46,1.71,1.94 \}$ is determined by imposing the constraint $H(0) = 0$ for each $y(0)=\{0,0.25,0.50,0.75,0.95 \}$.

It is important to note that in Figure \ref{fig:FuerzasdeMarea}, positive values of \(x\) are associated with one universe, whereas negative values of \(x\) correspond either to a distinct universe or to a separate spatial region within the same universe, and the geodesics that are able to traverse the wormhole are those that enter with an incidence angle close to the polar axis. In contrast, those geodesics with an initial angle close to the equatorial plane appear to be repelled by tidal forces and are consequently forced either to return or to be deflected toward the polar regions.
\begin{figure}[b]
    \centering
    \begin{minipage}{0.4\textwidth}
    \centering
        \includegraphics[width=\textwidth]{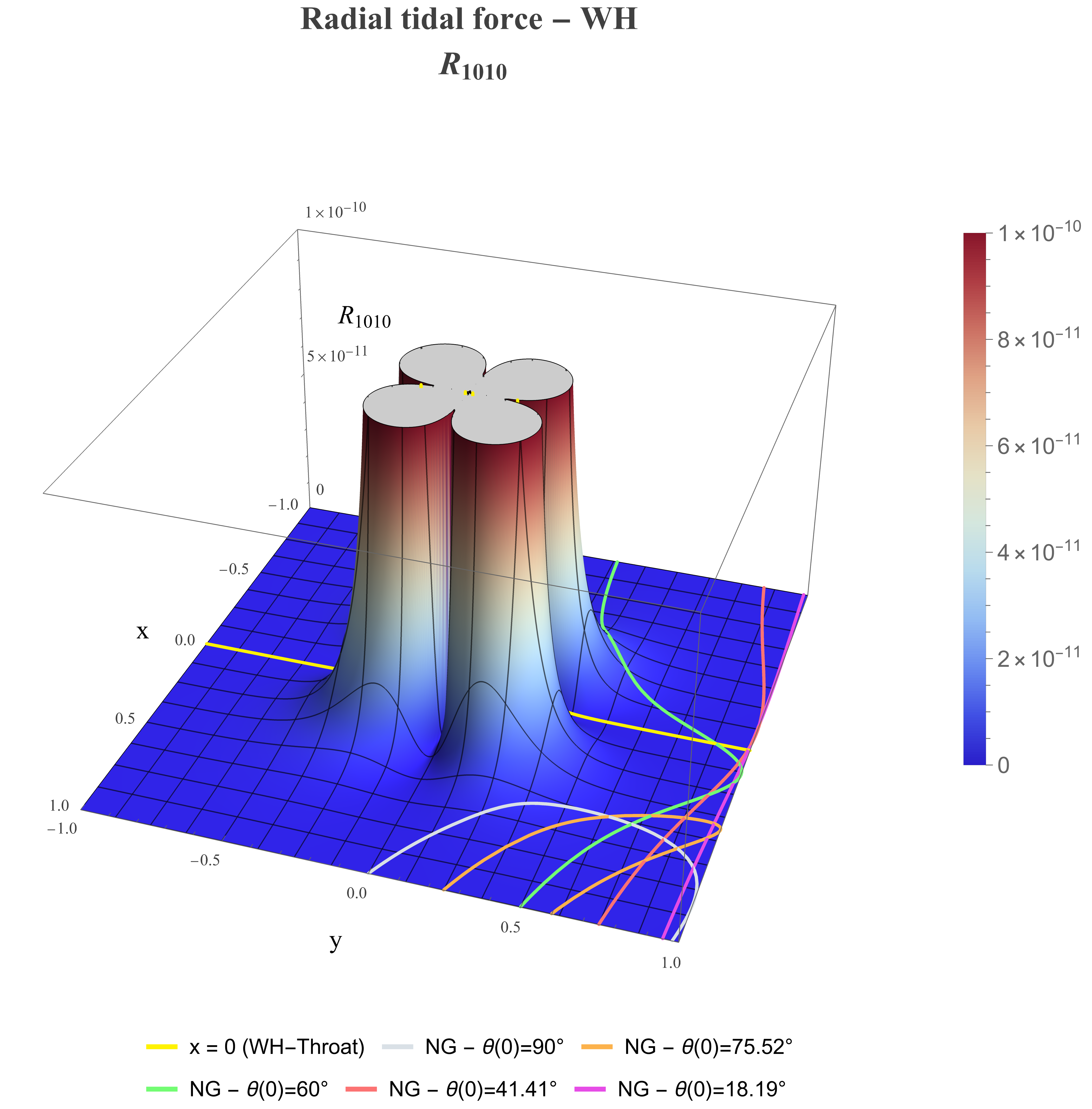}
        \subcaption{Radial tidal forces $[R_{1010}]=$ km$^{-2}$. The form and behavior are the same as those of the $\theta$-angular tidal force ($R_{2020}$).}
        \label{fig:TFRadial}
    \end{minipage}
    \hfill
    \begin{minipage}{0.4\textwidth}
    \centering
    \includegraphics[width=\textwidth]{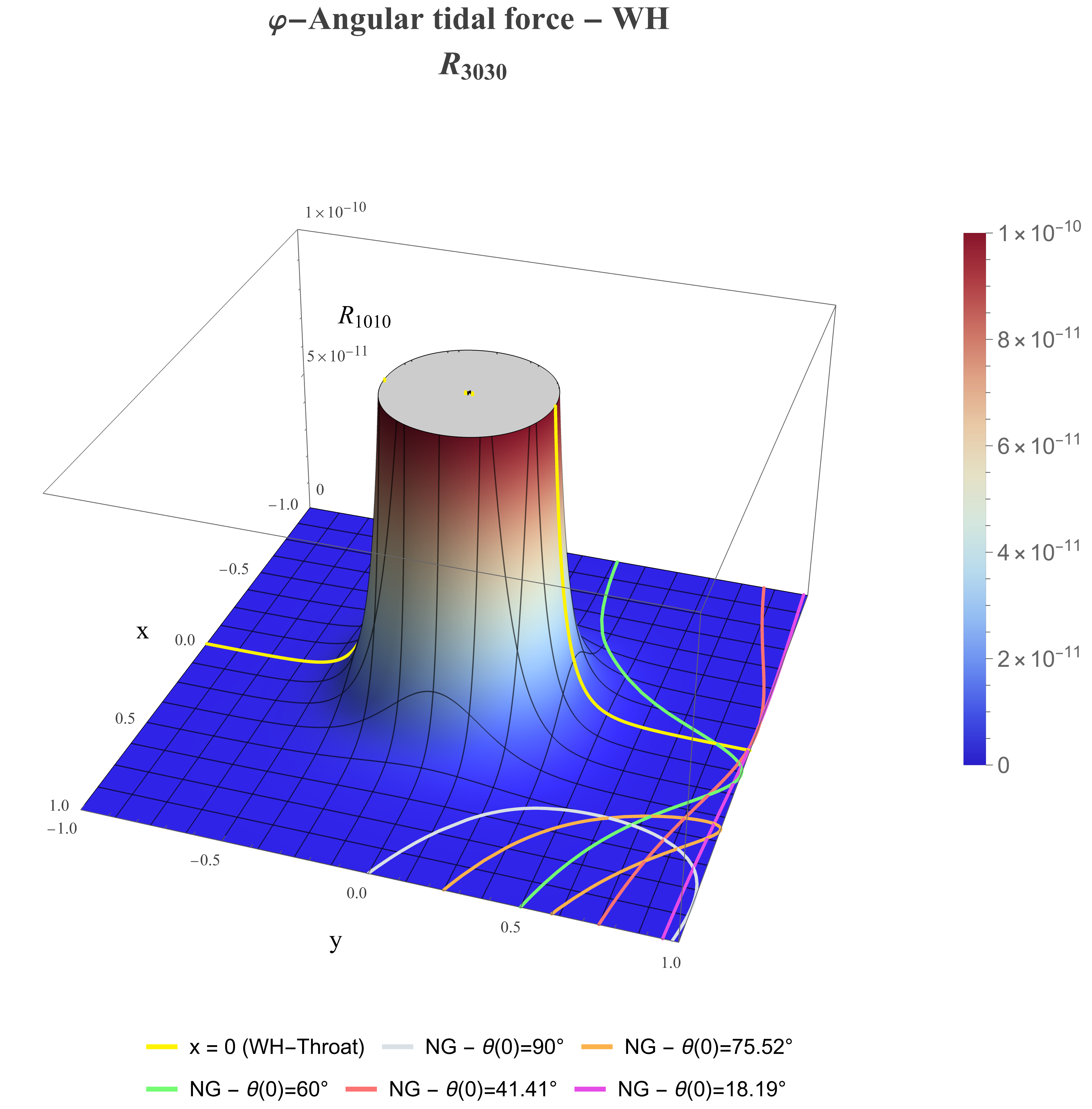}
        \subcaption{$\varphi$-Angular tidal forces $[R_{3030}]=$ km$^{-2}$}
        \label{fig:TFAngular}
    \end{minipage}
    \caption{Tidal forces corresponding to a traveler moving with non‑relativistic velocity. The parameters employed in the analysis were $\lambda_0 = 10^{-3}$, $\tau_0 = 10^{-4}$, $k_0 = 3/4$, and $L = 2$ km.In both figures, six continuous curves are plotted over the tidal surface. The yellow curve represents the tidal force evaluated at the throat of the wormhole, while the remaining five continuous curves correspond to null geodesics with distinct initial entry angles $\theta(0)$. }\label{fig:FuerzasdeMarea}
\end{figure}
By taking a Taylor series at the point $y=\pm 1$ in equation (\ref{Riemman en SR del astronauta - LambdaCombinada}), we obtain
{\small
{\setlength{\abovedisplayskip}{5pt}
 \setlength{\abovedisplayshortskip}{0pt}
\begin{align}\label{ComportamientoAsintotico}
        R_{1010}&=R_{3030}= R_{2020} \notag\\ 
        &\approx \frac{\Big(2\lambda_0x\pm \tau_0(x^2-1)\Big)^2}{4L^2 (x^2+1)^4}+\mathcal{O}(\varepsilon)^1.
    \end{align}
}
In this case, the upper sign refers to the Taylor series expanded about the point \(y = +1\), while the lower sign corresponds to the Taylor series expanded about the point \(y = -1\).
We observe that all components remain finite in the vicinity of the poles. In equation \eqref{ComportamientoAsintotico}, it is evident that the parameter $L$ plays a crucial role, as it sets the units of the tidal forces and determines the proportionality required to guaranty a physically acceptable (i.e., sufficiently small or safe) tidal force.


\section{Geodesics}
\subsection{Lagrangian and Hamiltonian formulation}

The Lagrangian and Hamiltonian formulations are crucial for studying particle dynamics. Using the metric, we can derive the corresponding Lagrangian $2\mathcal{L} \equiv g_{\mu \nu} \dot{x}^{\mu} \dot{x}^{\nu}$, so we have
\begin{multline}\label{Lagrangiano xy}
    2\mathcal{L} = -f(c\dot{t}-\omega \dot{\varphi})^2 +\frac{L_\pm ^2}{f} \bigg\{ (x^2\pm 1)(1-y^2) \dot{\varphi}^2 \\
    +e^{2k} (x^2\pm y^2) \bigg( \frac{\dot{x}^2 }{x^2\pm 1}+\frac{\dot{y}^2 }{1-y^2} \bigg)\bigg\},
\end{multline}
where $\dot{t} \equiv dt /ds$, $s$ is an affine parameter, and the upper sign corresponds to the WH, and the lower sign corresponds to the BH throughout this entire section.

From (\ref{Lagrangiano xy}), we can derive the moments using $p_{\mu} \equiv \partial \mathcal{L}/\partial x^{\mu}$. There are two cyclic coordinates ($t,\varphi$), therefore the constants of motion ($E,l_z$) are
\begin{subequations}\label{ConstantesMovimiento}
    \begin{equation}\label{Energia}
        -p_t= E=f(c\dot{t}-\omega \dot{\varphi}),
    \end{equation}
    \begin{equation}\label{Momento Angular}
        p_\varphi= l_z=\frac{L_{\pm}^2\, (x^2\pm 1)(1-y^2)}{f}\dot{\varphi}+\omega E.
    \end{equation}
\end{subequations}
Isolating $\dot{t}$ and $\dot{\varphi}$ from the equations (\ref{ConstantesMovimiento}), we can divide the two to derive the proper angular velocity
\begin{equation}\label{Velocidad Angular}
     \frac{ \dot{\varphi} }{ \dot{t} } =\frac{d \varphi}{dt}=\frac{f^2 (l_z-\omega E) c}{L_\pm ^2(x^2\pm 1)(1-y^2)E +f^2 \omega (l_z-\omega E)},
\end{equation}
The other two moments are determined by
\begin{subequations}\label{Momentos px py}
    \begin{equation}\label{px}
        p_x= \frac{L^2(x^2\pm y^2) e^{2k}}{f(x^2\pm1)}\dot{x},
    \end{equation}
    \begin{equation}\label{py}
        p_y= \frac{L^2(x^2\pm y^2) e^{2k}}{f(1-y^2)}\dot{y}.
    \end{equation}
\end{subequations}
To obtain the Hamiltonian, it is necessary to evaluate $2\mathcal{H} \equiv g^{\mu \nu} p_\mu p_\nu$. Considering all momentum components and the inverse metric (\ref{ds sp}), the corresponding Hamiltonian can be determined as
\begin{multline}\label{Hamiltoniano xy}
    2\mathcal{H} = \frac{f \big( l_z -\omega E\big)^2}{L_\pm^2(x^2\pm1)(1-y^2)}-\frac{E^2}{f} \\
    +\frac{f e^{-2k}}{L_\pm ^2(x^2\pm y^2)} \bigg\{ (x^2\pm 1) p_x^2 + (1-y^2) p_y^2 \bigg\}.
\end{multline}
%


From the expressions for the moment, we obtain the moment norm, which is given by
\begin{equation*}
    g^{\mu \nu}p_{\mu}p_{\nu}=(m_0)^2 c^2 =2 \mathcal{H},
\end{equation*}
where $(m_0)$ represents the rest mass, that is, for photons we have $(m_0)=0$. Therefore, to determine the initial conditions corresponding to the null geodesics, it is necessary to impose the condition $\mathcal{H} (x^\mu , p^\mu)=0$ to (\ref{Hamiltoniano xy}).

Employing the solution presented in (\ref{SolucionLambdaCombinada}), together with the values
\begin{subequations}\label{CondicionesIniciales GeodesicNull}
\begin{equation}\label{Condiciones de E lz}
    l_z=-5, \qquad E=10,
\end{equation}
\begin{equation}
    x(0)=10, \qquad p_y(0)=1,
\end{equation}
\end{subequations}
for an arbitrary choice of $y(0)=\{ 0,0.25,0.50,0.75,0.92 \}$, we find $-p_x(0) \approx \{19.89,19.90,19.92,19.94,19.93 \}$, in such a way that $\mathcal{H}\big(x^\mu(0), p^\mu(0)\big) = 0$ holds.

\subsection{Null geodesic graphics}

The equations of motion are derived from Hamilton's equations given by
{\setlength{\abovedisplayskip}{0pt}
 \setlength{\abovedisplayshortskip}{0pt}
    \begin{equation}\label{Ecuaciones de Hamilton}
        \dot{x}^{\mu} \equiv \frac{\partial \mathcal{H}}{\partial p_{\mu}}, \qquad \qquad \dot{p}_{\mu} \equiv -\frac{\partial \mathcal{H}}{\partial x^{\mu}}.
    \end{equation}
}
Using the initial condition (\ref{CondicionesIniciales GeodesicNull}), which specifies the null geodesics, together with the corresponding equation of motion (\ref{Ecuaciones de Hamilton}), we obtain a numerical solution from which we can generate plots of $x(s)$, $y(s)$, $\varphi(s)$, and $t(s)$. These plots allow us to examine how the null geodesic behaves as it either passes through the WH or avoids reaching the causally disconnected singularity in the SU-E scenario. For the S-E scenario, we can observe the behavior of the geodesic as it does not intersect the event horizon.
In Figure \ref{fig:GdeosicasSpheroidalesWH}, null geodesics corresponding to various initial conditions $y_0$ are depicted in different colors, taking into account the WH. The plots use oblate spheroidal coordinates to better display the ring singularity, depicted as a gray dot, and the WH throat is depicted by a fuchsia dotted line. The region $x>0$ corresponds to one universe, whereas $x<0$ represents either a different universe or another region of the same universe. One observes that the null geodesic in the equatorial plane ($y_0=0$) is unable to cross the WH. In contrast, all other null geodesics traverse the WH without obstruction.

On the other hand, focusing on the BH case, Figure \ref{fig:GdeosicasSpheroidalesBH} displays the null geodesics for various initial conditions $y_0$, each indicated by a different color. The plots are presented in oblate spheroidal coordinates to more clearly exhibit the ring singularity, shown as a gray dot. The event and Cauchy horizons are represented by dotted fuchsia lines, while the surface singularities are indicated by gray dot-dashed lines. This figure was generated using the same initial conditions and parameter values as in the WH case. In other words, Figures \ref{fig:GdeosicasSpheroidalesBH} and \ref{fig:GdeosicasSpheroidalesWH} were obtained employing identical parameters and initial conditions.

Figures \ref{fig:GeodesicasNulasWH} and \ref{fig:GeodesicasNulasBH} show the photon trajectories in Cartesian coordinates for the WH and BH cases, respectively. In both figures, the ring singularity is depicted as a fuchsia torus with radius $r = l_1$ on the equatorial plane. For the WH configuration, a yellow $\mathbb{S}^2$ sphere represents the wormhole throat; solid curves correspond to universe 1, while dashed curves correspond to universe 2. In the BH configuration, the outer black $\mathbb{S}^2$ sphere denotes the event horizon and the inner black $\mathbb{S}^2$ sphere denotes the Cauchy horizon. In this case, the surface singularities are illustrated by yellow surfaces, and only solid trajectories in different colors are used.

To derive the null geodesics in Cartesian coordinates, we used:
\begin{align*}
    X_c&=(L_\pm x(s) +l_1) \sqrt{1-y(s)^2} \,\cos{\Big(\varphi (s)\Big)} ,\\
    Y_c&=(L_\pm x(s) +l_1) \sqrt{1-y(s)^2} \,\sin{\Big(\varphi (s)\Big)} ,\\
    Z_c&=(L_\pm x(s) +l_1) y(s).
\end{align*}
\begin{figure*}
    \begin{minipage}{0.49\textwidth}
        \centering
        \begin{subfigure}{\textwidth}
            \includegraphics[width=\textwidth]{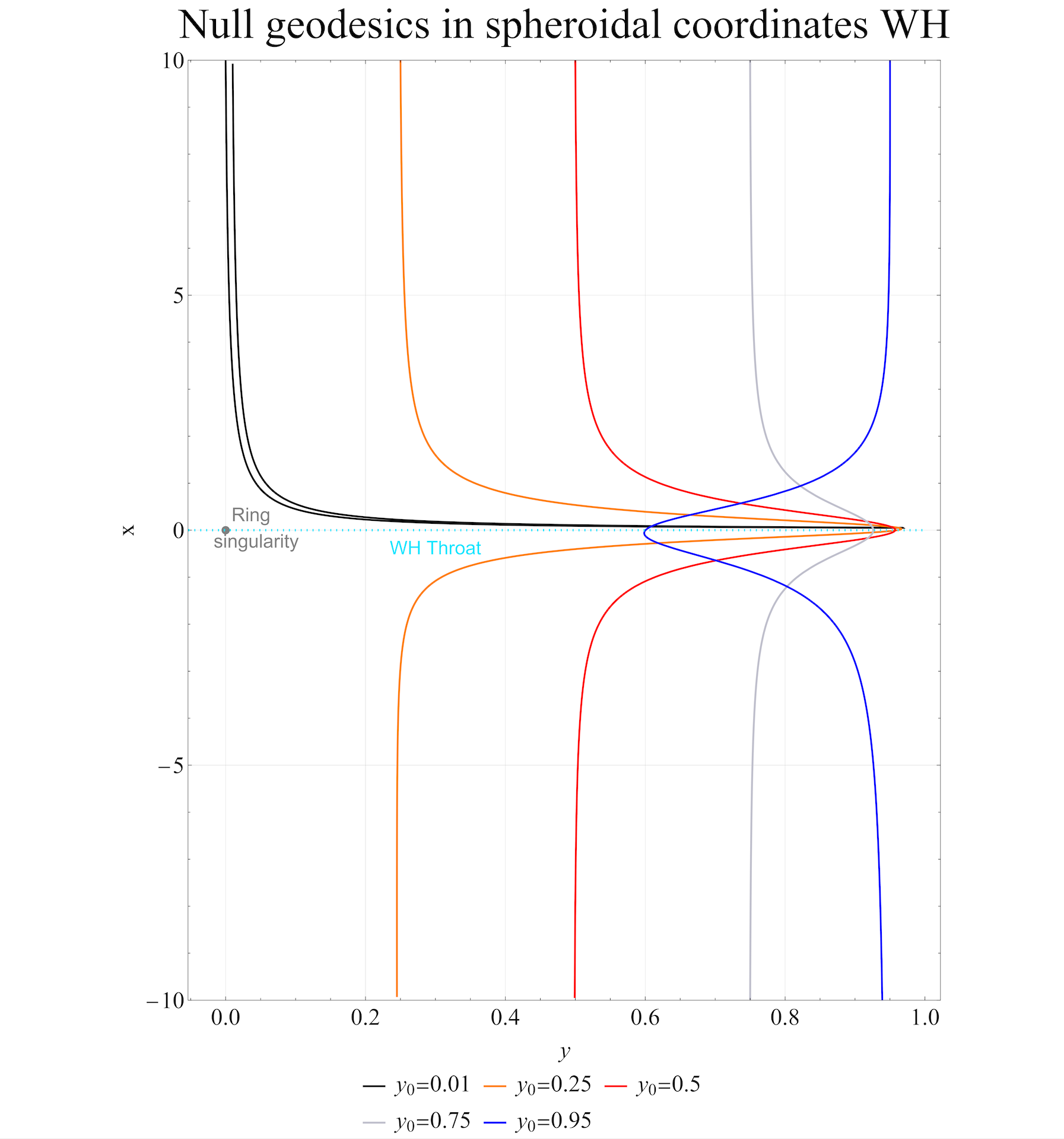}
            \caption{Null geodesics for different values of $y_0$ in spheroidal oblate coordinates. Positive values of $x$ correspond to one universe, while negative values of $x$ pertain to either another universe or the same universe in a distinct spatial region.}
            \label{fig:GdeosicasSpheroidalesWH}
        \end{subfigure}
    \end{minipage}%
    \hfill
    \begin{minipage}{0.49\textwidth}
        \centering
        \begin{subfigure}{\textwidth}
            \includegraphics[width=\textwidth]{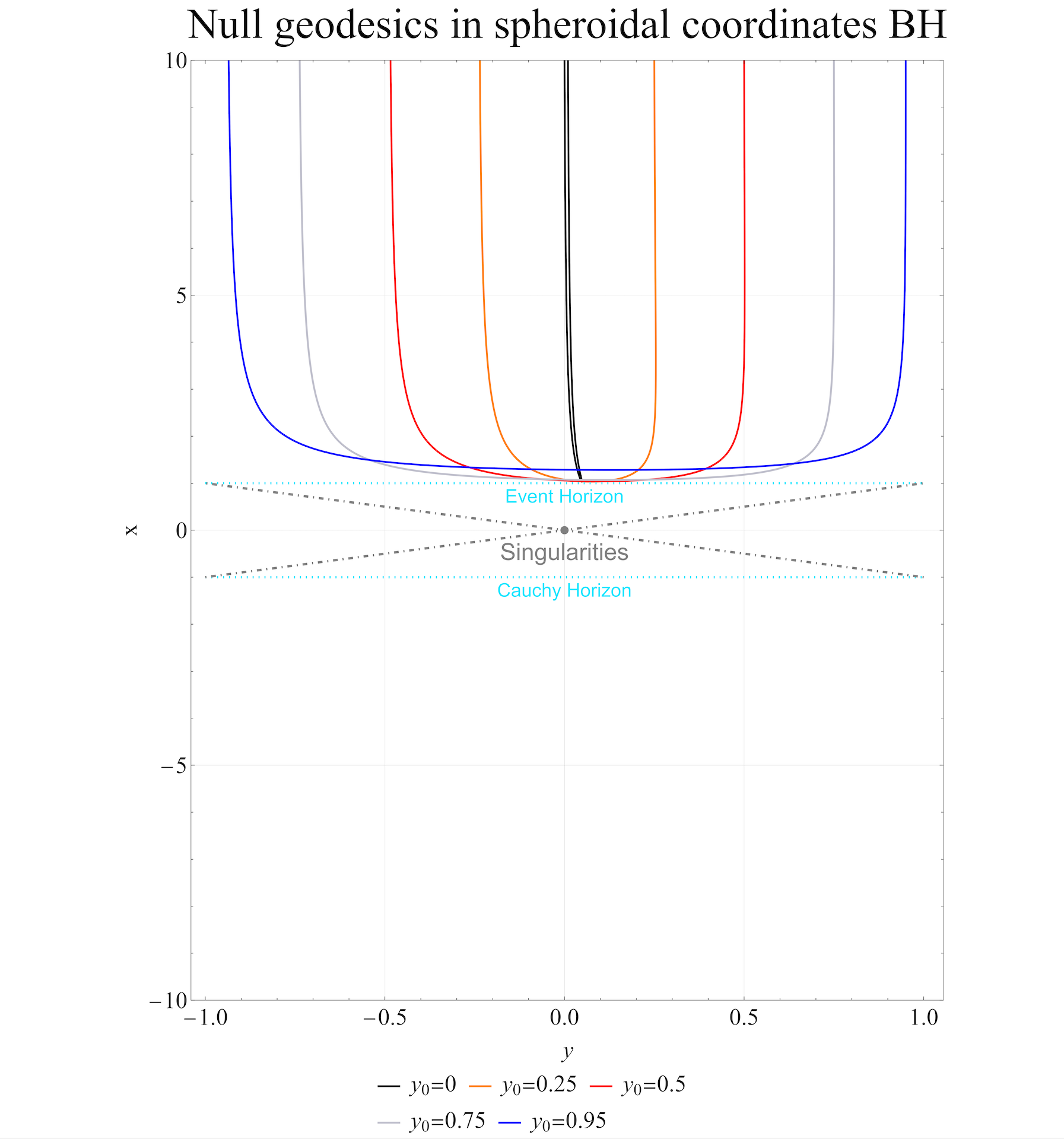}
            \caption{Null geodesics for different values of $y_0$ in spheroidal prolate coordinates. The surface singularities are indicated by the gray dashed-point lines, while the ring singularities are represented by gray dot.}
            \label{fig:GdeosicasSpheroidalesBH}
        \end{subfigure}
    \end{minipage}
    \centering
    \begin{minipage}{0.49\textwidth}
        \centering
        \begin{subfigure}{\textwidth}
            \includegraphics[width=\textwidth]{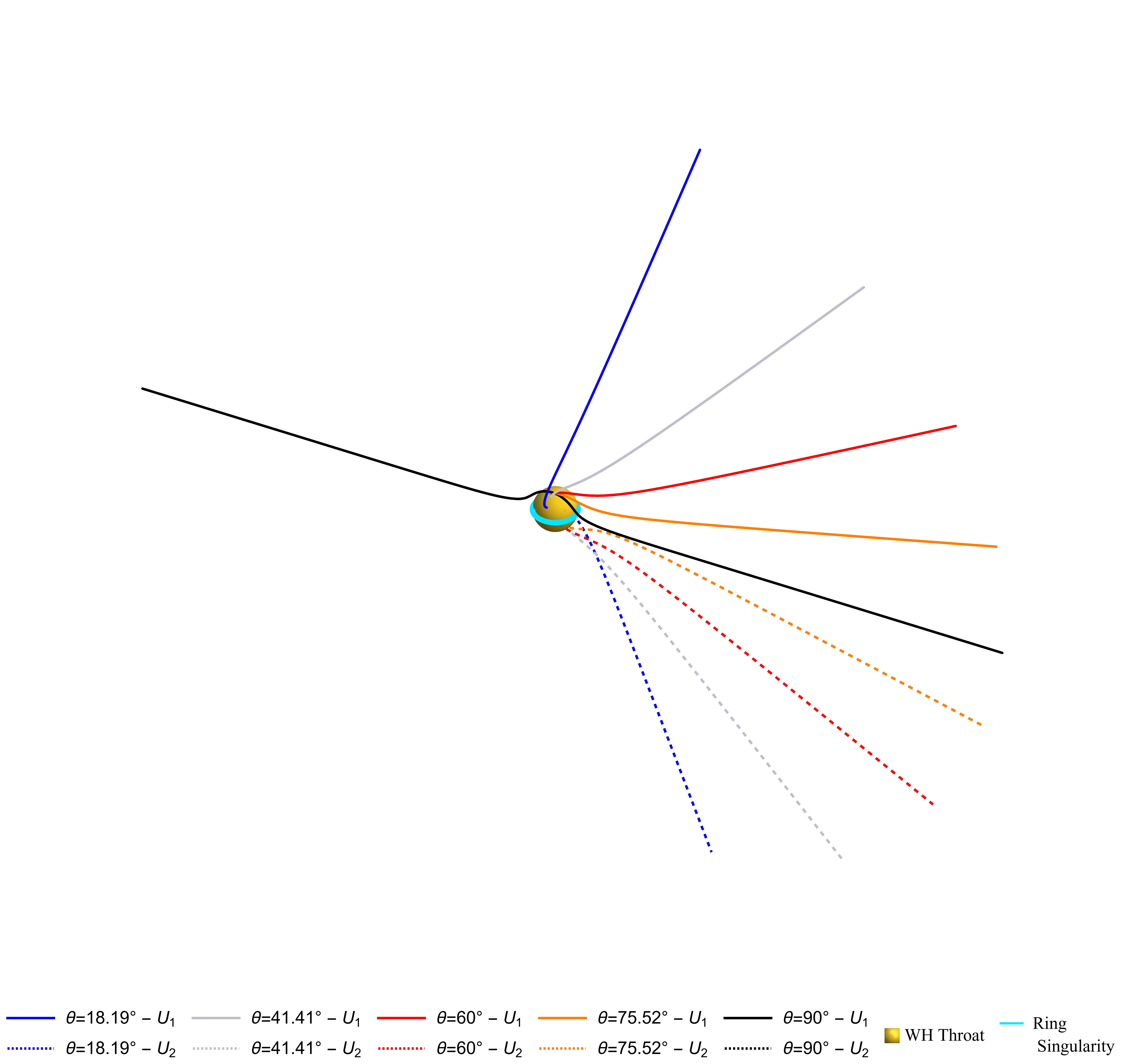}
            \caption{Schematic depiction of the null geodesics in Cartesian coordinates and the wormhole. The dotted lines represent the second universe, while the continuous lines represent the first universe. The yellow sphere indicates the wormhole throat, and the fuchsia torus corresponds to the ring singularity.}
            \label{fig:GeodesicasNulasWH}
        \end{subfigure}
    \end{minipage}%
    \hfill
    \begin{minipage}{0.49\textwidth}
        \centering
        \begin{subfigure}{\textwidth}
            \includegraphics[width=\textwidth]{GeodesicasNulasBH.png}
            \caption{Schematic illustration of the null geodesics and the internal structure of the black hole. The outer black sphere represents the event horizon, while the inner black sphere corresponds to the Cauchy horizon. The fuchsia torus denotes the ring singularity, and the yellow surface indicates the surface singularities.} \label{fig:GeodesicasNulasBH}
        \end{subfigure}
    \end{minipage}
    \caption{ Null geodesic curves that satisfies $\mathcal{H} \big(x^\mu (0), p^\mu(0)\big)=0$ and with parameters $L_\pm=2$, $l_1=1$, $\lambda_0=10^{-3}$, $\tau_0=10^{-4}$, $k_0=3/4$.}
    \label{fig:Geodesicas}
\end{figure*}
%
%
%

To find the proper time and the effective potential of a null geodesic, we begin with the Lagrangian (\ref{Lagrangiano xy}) written in Weyl coordinates $(\rho,z)$.
\begin{equation}\label{Lagrangiano rhoz}
    2\mathcal{L} = \frac{e^{2k}}{f}(\dot{\rho}^2+\dot{z}^2)-\frac{1}{f}\bigg(E^2-\frac{f^2(l_z-\omega E)^2}{\rho^2} \bigg),
\end{equation}
So, if we set $2\mathcal{L}=0$, we get the equations of motion for the photons, and we can see the shape of the effective potential
{\small}
{\setlength{\abovedisplayskip}{0pt}
 \setlength{\abovedisplayshortskip}{0pt}
\begin{align}\label{Ecuacion de movimiento}
    e^{2k}(\dot{\rho}^2+\dot{z}^2)=& E^2- \frac{f^2(l_z-E \omega )^2}{\rho^2}\notag\\
    =&E^2\bigg(1-\frac{f^2(b-\omega )^2}{\rho^2} \bigg),
\end{align}
}}
where $f>1$ for the interest region, note that $\dot{\rho}^2 + \dot{z}^2 > 0 \quad \forall s \in \mathbb{R}$, the exponential term fulfils $e^{2k} > 0 \quad \forall s \in \mathbb{R}$, the impact parameter is $b=l_z/E$ and 
\begin{equation}\label{potencial efectivo}
    V_{eff}=\frac{f^2(b-\omega )^2}{\rho^2},
\end{equation}
so that $V_{eff}\leq 1$ is satisfied for null geodesics to exist.

Let us consider the case in which the throat in the equatorial plane is not located at $x_G \approx 0.0293$, but instead is positioned at $x = 0$. If we then launch a null geodesic toward the ring singularity in the equatorial plane and compute the proper time required for it to reach the singularity, this proper time is given by
\begin{equation}\label{IntegralTiempoPropio}
    s=\int_{x_0}^{x_1} \frac{e^{k(x,y_0)} d\Delta} {\sqrt{E^2-f^2(l_z-\omega(x,y_0) E)^2/\rho(x,y_0)^2}}
\end{equation}
where 
{\setlength{\abovedisplayskip}{-2pt}
 \setlength{\abovedisplayshortskip}{-2pt}
\begin{equation*}
    d\Delta=\sqrt{(d\rho^2+dz^2)_{|y:cte}}=L\sqrt{\frac{x^2+y_0^2}{x^2+1}} \quad dx.
\end{equation*}
}
By integrating expression (\ref{IntegralTiempoPropio}) from the initial condition $x_0 = 1$ and plotting the function $-s(x_1)$ in the limit $x_1 \to 0$, while incorporating the parameters specified in (\ref{Condiciones de E lz}), Figure \ref{fig:TiempoPropio} shows that the proper time required for the null geodesic to reach the ring singularity diverges as $x_1$ approaches zero. More importantly, the numerical integration reveals that, before the geodesic can reach the singularity, it first intersects the wormhole throat $x_G$ and continues into the other universe. This behavior provides further evidence that the ring singularity is causally disconnected.
\begin{figure}[b]
    \centering
    \includegraphics[width=0.8\linewidth]{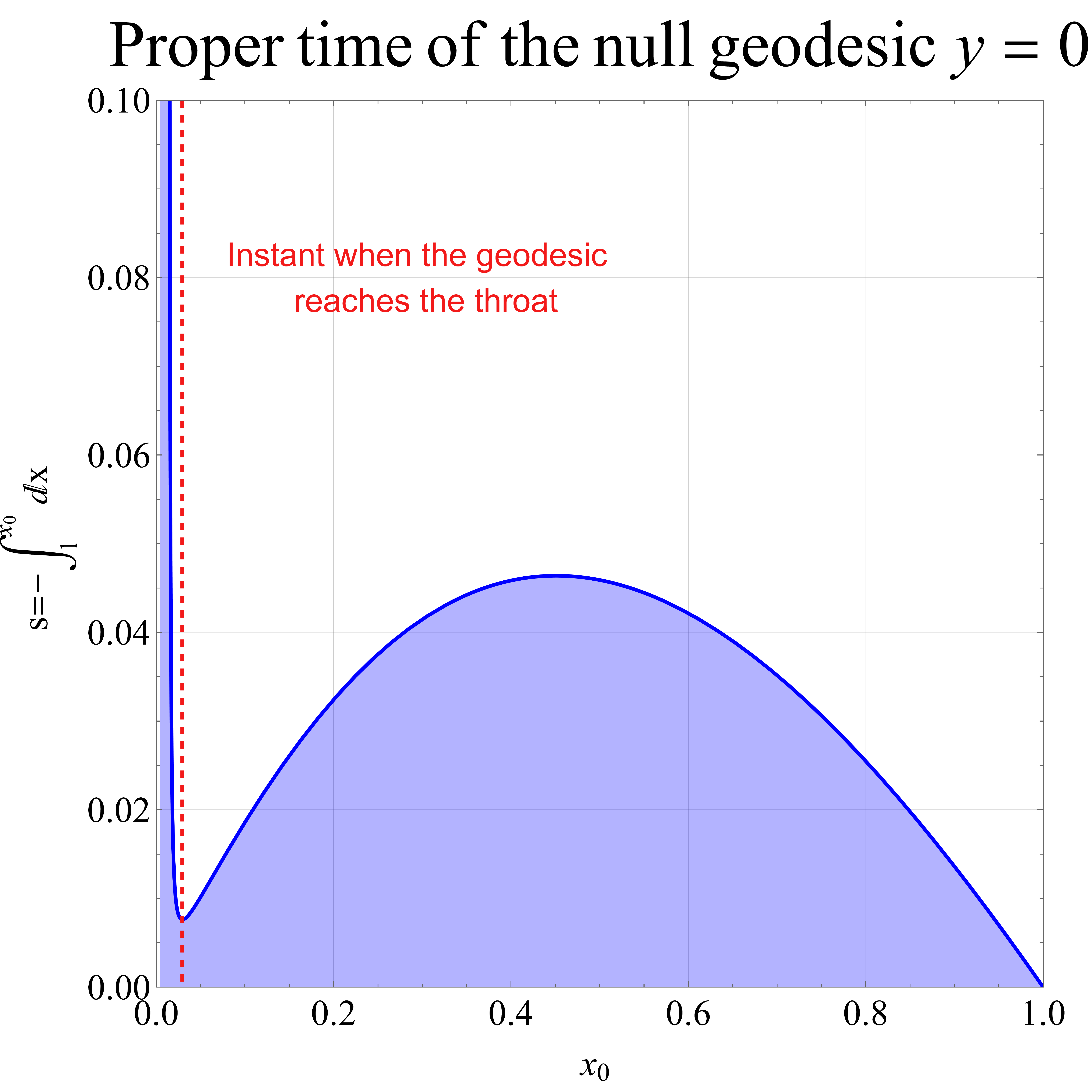}
    \caption{Value of the integral (\ref{IntegralTiempoPropio}) computed with the parameters $\lambda_0=10^{-3}$, $\tau_0=10^{-4}$, $L=2$, and $k_0=3/4$, represented by the blue curve. The dashed red line indicates the exact instant at which the geodesic reaches the wormhole throat and then diverges to infinity as $x_0$ tends to zero.}
    \label{fig:TiempoPropio}
\end{figure}
%
The effective potential of the photons is \eqref{potencial efectivo} and its minimal unstable points are identified by
\begin{equation*}
    \frac{d}{dx}V_{eff}=0,\qquad \qquad \frac{d^2}{dx^2}V_{eff}<0.
\end{equation*}
Figure \ref{fig:PotencialEfectivo} shows the profile of the effective potential, where we compare three different values of $y$. The plot reveals the presence of two unstable minima. The most representative curve corresponds to $y_0 = 0.25$, and for this case we have indicated the associated minima at approximately $x_{p2} \approx -0.2152$ and $x_{p1} \approx 0.2610$.
\begin{figure}[b]
    \centering
    \includegraphics[width=0.8\linewidth]{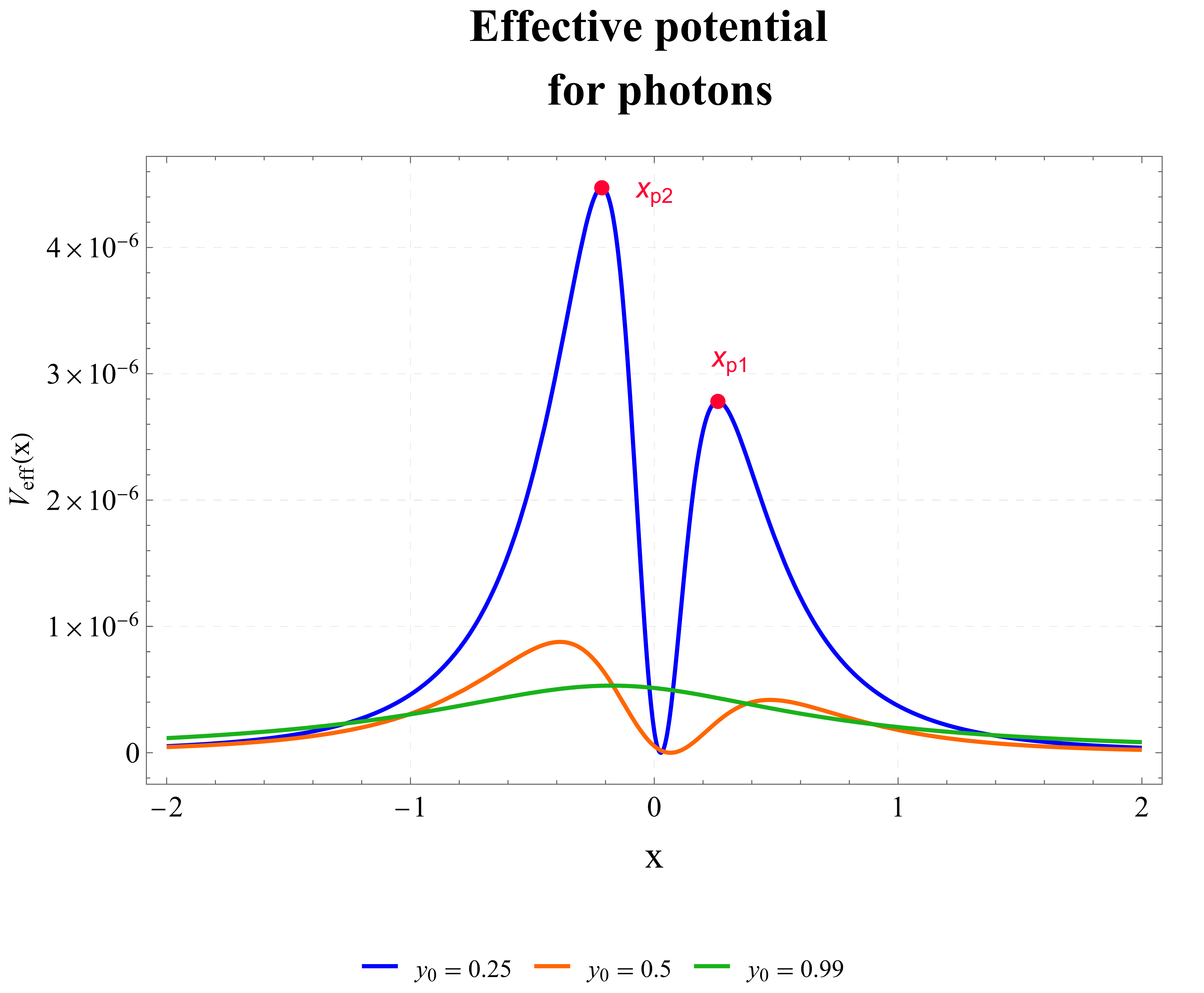}
    \caption{Effective potential \eqref{potencial efectivo} computed using $\lambda_0 = 10^{-3}$, $\tau_0 = 10^{-4}$, $L = 2$, $l_z = 0$, and different values of $y$. The red points correspond to the minimally unstable points of the effective potential for $y = 0.25$.}
    \label{fig:PotencialEfectivo}
\end{figure}

We will use the following values as the basis for our analysis.

\paragraph{Sun.} 
The solar mass $M_\odot$ is obtained from standard heliophysical data (NASA fact sheets and IAU nominal solar parameters), giving a Schwarzschild radius $r_{sc\odot}/2 = 1.4766 \times 10^3$m. The solar angular momentum, inferred from helioseismology, is $J_\odot \simeq 1.92 × 10^{41}$kg m$^2$ s$^{-1}$, from which we obtain $J_\infty = 4.7560 \times 10^5$m$^2$.

\paragraph{Neutron star (canonical $1.4 M_\odot$).} 
For neutron stars with undetermined masses, we adopt the standard value $M = 1.4\,M_\odot$, consistent with population analyses and observations of binary pulsars, resulting in $r_s/2 = 2.0673 \times 10^{3}$m. For the rapidly rotating pulsar with an observed spin frequency $f = 716$Hz, we have $\Omega = 2\pi f = 4498.7607$rad/s. Using the representative moment of inertia $I_{\mathrm{mid}} = 1.6107\times 10^{38}$kg m$^2$, the angular momentum is $J = I_{\mathrm{mid}}\,\Omega = 7.2460\times 10^{41}$kg m$^2$ s, which in geometric units is $J_{\infty}= \frac{G}{c^3}\,J= 1.7949\times 10^{6}$m$^2$.
Following a conservative approach based on orbital-dynamics constraints, we adopt an upper bound on the net electric charge of a neutron star $|Q|\lesssim 5\times10^{19}\,\mathrm{C}$, then $Q_L=\sqrt{\frac{G}{4\pi\varepsilon_0}}\frac{Q}{c^2}
\lesssim 4.31\times10^{2}\,\mathrm{m}$, as inferred from binary pulsar phenomenology within an effective Reissner-Nordstrom description \cite{Iorio:2012dbo}.

\paragraph{Magnetar.}
For magnetars, the rotation period $P$ is directly observable. We adopt $P = 3.24\,\mathrm{s}$, consistent with the values given in magnetar timing catalogs. Therefore, the angular momentum is $J = I\,\Omega = I\,\frac{2\pi}{P}$, where $I$ is the stellar moment of inertia. Since $I$ depends on the assumed equation of state, we employ a commonly used representative value for the neutron star's moment of inertia, $I_{\mathrm{mid}}$, which results in $J_\infty = 7.7370\times 10^{2}$m$^2$.

\paragraph{Sagittarius A*.}
For the supermassive black hole at the galactic center, the stellar orbital dynamics give a mass of $M = 4.152 \times 10^6 M_\odot$, with a Schwarzschild radius $r_{s}/2 = 6.1309 \times 10^9$m. The dimensionless spin is $\chi \equiv cJ/(GM²)$; for $\chi = 0.9$, we obtain $J_\infty = \chi\, l_1^2 = 3.3832 \times 10^{19}$m$^2$.

In Table \ref{ValoresDeParametrosReferenciasReales}, we summarize the above values and add the corresponding parameter $a$. This table serves as a reference for real-world conditions, while to determine the dimensional expressions associated with the parameters of our compact objects we use the Eqs. \eqref{Super-Extreme}, \eqref{eq:NUT_dualKomarInfinito}, \eqref{eq:MomentoAngular_KomarInfinito}, \eqref{eq:Q_KomarInfinito}, \eqref{eq:H_Komar} and \eqref{eq:H_KomarInfinito}, with the Table \ref{ValoresDeParametrosWH} presenting the corresponding parameter values for wormholes (SU-E), for objects analogous to those discussed above with the addition of a row for an Earth-sized object, and the Table \ref{ValoresDeParametrosBH} presenting the corresponding values for a black hole (S-E).
%
%
\section{Electromagnetic field}

To visualize the detailed structure of the electromagnetic field given in (\ref{CampoElectromagneticoxy}), we employ the metric functions defined in (\ref{SolucionLambdaCombinada}). This procedure enables a schematic representation of the resulting field configuration. For the corresponding plot, the parameters are chosen as $\lambda_0 = 10^{-3}$, $\tau_0 = 10^{-4}$, and $c = 1$, with $\sigma_0 = 1$, $L = 2$, and $l_1 = 1$.

Several important observations can be drawn from equations \eqref{CampoElectromagneticoxy}. First, on the equatorial plane $y=0$, the electromagnetic vector field remains fully regular and finite for $x\neq 0$, while at the ring singularity, characterized by $x=0$ and $y=0$, the electromagnetic field is no longer defined. Likewise, on the throat $x=0$, the electromagnetic field is regular and finite everywhere except at the point $y=0$.

Figure \ref{fig:CampoElectromagnetico} illustrates the vector fields corresponding to the magnetic and electric fields in spheroidal oblates coordinates (WH), shown in the first and second rows, respectively.

For both panels of Fig.~\ref{fig:CampoElectromagnetico}, the ring singularity is located at the point $x=0=y$. It should be emphasized that all plots are schematic and do not correspond to actual values in SI units. The essential feature is that, in the regions adjacent to the ring singularity, the electromagnetic field is extremely intense and becomes significant in the inner domain of the white circle formed by the white vectors, namely the region delineated by the red row vectors. At the ring singularity itself, the electromagnetic field diverges, tending to infinity.

On the other hand, we plot the geodesic curves using the same initial conditions and parameters employed to generate Fig.~\ref{fig:FuerzasdeMarea}. With this in mind, we observe that all geodesics avoid the high–field regions of the electromagnetic field (the inner region of the white circle formed by the white arrows), which results in an effective repulsive behavior. The most illustrative case is the geodesic launched from the equatorial plane, $y(0)=0$, which is deflected to the right and is unable to traverse the wormhole, that is, it cannot cross from regions with $x>0$ to regions with $x<0$. The wormhole throat is represented by the blue curve. 

In contrast, when a geodesic is initiated near the poles, e.g. at $y=1$, it is able to propagate from the region with $x>0$ to that with $x<0$, avoiding the domain in which the electromagnetic field attains its largest magnitudes.
\begin{figure*}
    \begin{minipage}{0.6\textwidth}
        \centering
        \begin{subfigure}{\textwidth}
            \includegraphics[width=\textwidth]{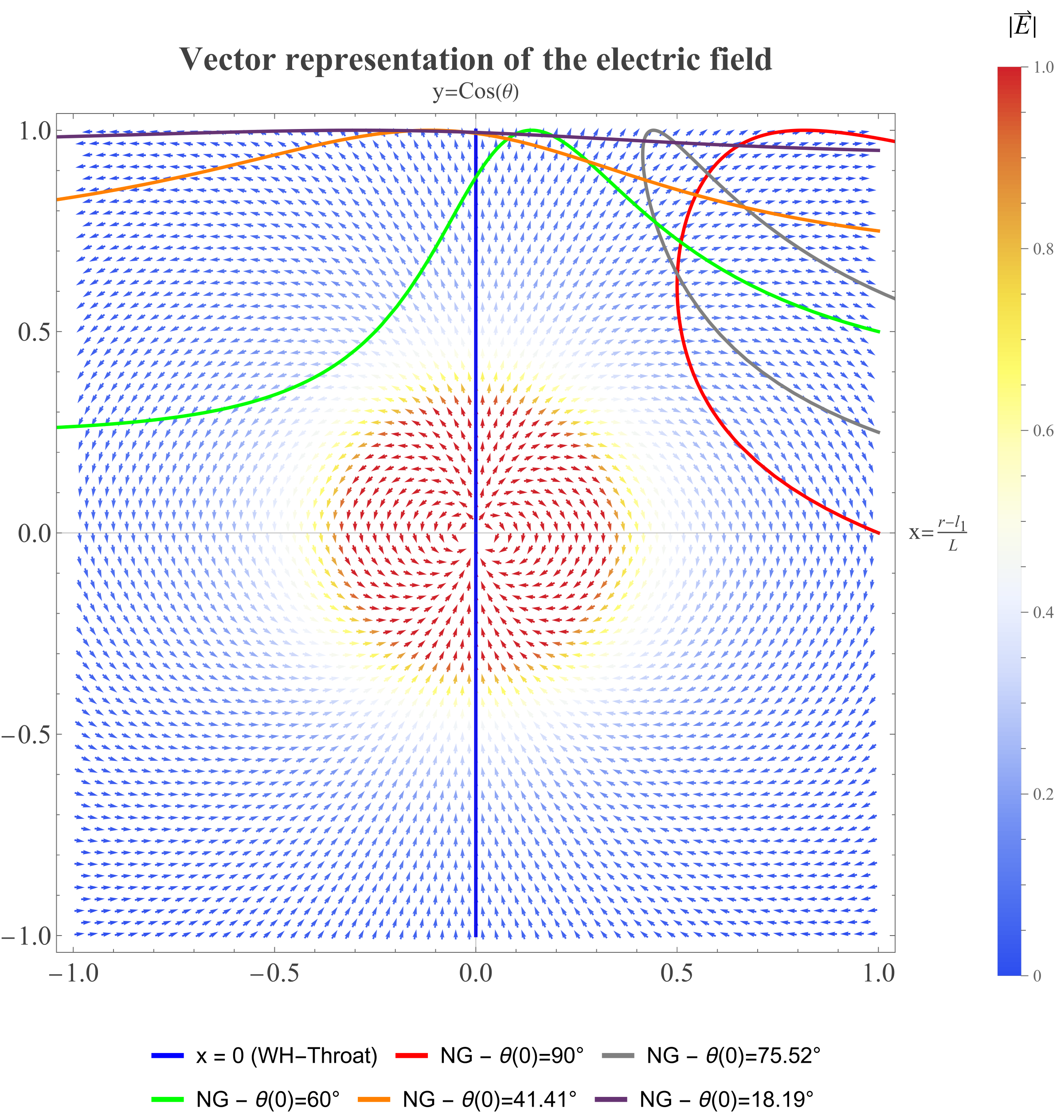}
            \caption{}
            \label{fig:GraficaCampoElectrico}
        \end{subfigure}
    \end{minipage}%
    \hfill
    \begin{minipage}{0.6\textwidth}
        \centering
        \begin{subfigure}{\textwidth}
            \includegraphics[width=\textwidth]{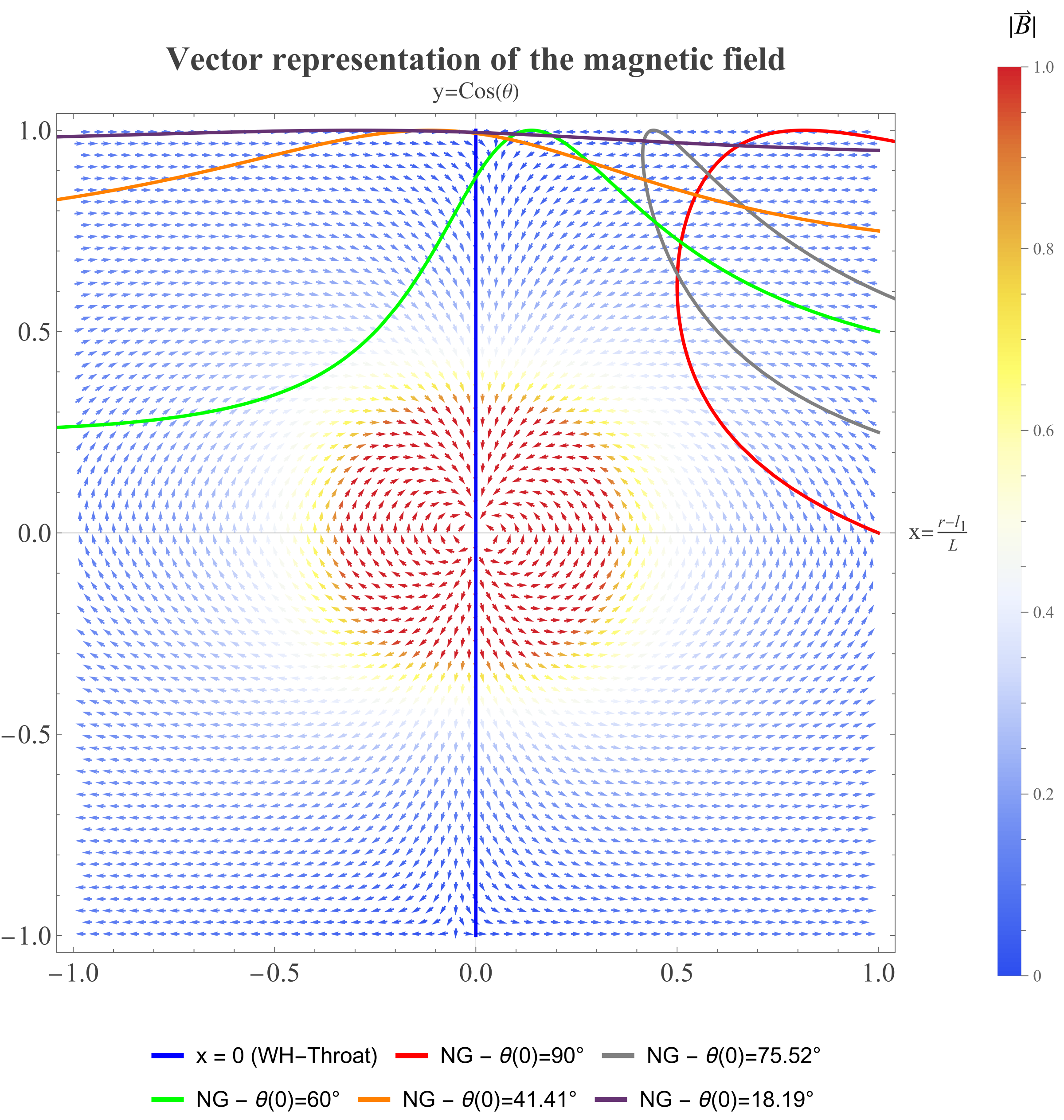}
            \caption{}
            \label{fig:GraficaCampoMagnetico}
        \end{subfigure}
    \end{minipage}%
    
    \caption{Diagrammatic electromagnetic vector field. Positive values of \(x\) are associated with one universe, whereas negative values of \(x\) correspond either to a distinct universe or to a separate spatial region within the same universe. }
    \label{fig:CampoElectromagnetico}
\end{figure*}
%


\section{Conclusions}

In this work, using the definition of invariant conserved charges, we show that our exact solution corresponds to a genuinely dionic configuration, with nonzero NUT charge $N_\infty$, electric charge $Q_\infty$, and magnetic charge $H_\infty$, all fixed by a single parameter, $\tau_0 \, L_{\pm}/2$. The configuration has zero Komar mass, $M_\infty = 0$, but nonzero Komar angular momentum $J = - (L_\pm)^2 \lambda_0 / 2 f_0$. If $\tau_0\neq 0$ it is embedded in a NUT-type spacetime, characterized by the asymptotic gravitomagnetic contribution $-(L_\pm)\,\tau_0 \cos\theta$ and the rotational frame-dragging term $(L_\pm)^2 \lambda_0 \sin^2\theta / r$. If $\tau_0=0$ the solution represents an electromagnetic dipole \cite{DelAguila:2015isj}.
These properties are established by analysing the asymptotic behaviour of the metric functions and the electromagnetic four-potential, as well as by computing and examining the five Newman–Penrose Weyl scalars and the Newman–Penrose Maxwell scalars constructed from an appropriate Newman–Penrose null tetrad. In addition, we demonstrate that our new exact black hole solution (sub-extremal regime) is consistent with the Cosmic Censorship Conjecture, namely, all curvature singularities and CTCs are confined to the interior region of the event horizon. The spacetime exhibits the same ring singularity as in the WH configuration, but now supplemented by two additional surface singularities. Furthermore, we analyse the mechanism by which the WCCC operates in this context. In particular, we explain how the WH throat is shifted to the value $x_G$ such that the condition $0 < |x_v| < |x_G|$ is satisfied, indicating that the throat effectively hides all causal pathologies of spacetime (singularities and CTCs), using the Areal function invariant.

Concerning the NEC, our findings are consistent with those reported in \cite{DelAguila:2015isj} and \cite{Bixano:2025jwm}. In particular, for dilatonic wormholes, the NEC is fulfilled. More significantly, we demonstrate that in the black hole configuration the NEC is always satisfied in the entire region exterior to the event horizon. Furthermore, at the event horizon itself, the NEC continues to hold. By contrast, in the interior region, it is not possible to establish the validity of the NEC for $|x<1|$.

Using Figures \ref{fig:FuerzasdeMarea} and \ref{fig:CampoElectromagnetico}, it is found that geodesics passing through the WH are consistently deflected toward regions near the polar axis. This behavior appears to arise from their tendency to avoid regions characterized by large magnitudes of tidal forces and strong electromagnetic fields. In contrast, geodesics that start close to the equatorial plane are reflected and stay within the same universe, thereby being effectively blocked from traversing the WH.

Regarding the temporal behavior associated with the naked singularity, if the wormhole throat is omitted, the coordinate time required for a geodesic to reach the singularity diverges, implying that this naked singularity is causally disconnected. When the throat is taken into account, the geodesic first traverses to the opposite universe through the WH before encountering the naked singularity. For the WH effective asymmetry potential, two unstable equilibrium points are present, one in each region. These unstable points give rise to photon spheres, providing a natural criterion for distinguishing wormholes from black holes based on their geodesic structure (See \cite{Wielgus:2020uqz}, \cite{Vincent:2020dij}).

Analyzing the WH geometry, it is found that, for trajectories confined to the equatorial plane, the minimum value of the throat radius approaches the quantity \(L_+\) in the embedding diagram, and the corresponding shape throat tends to close. In this regime, the WH throat also encloses the ring singularity. 

In summary, the optimal entry angle that avoids encountering significant impediments is located in regions close to the polar axis. This follows from the fact that, in these regions, both tidal forces and electromagnetic fields attain minimal and finite values.

To establish a lower bound for evaluating the safe size of a WH, an analysis based solely on tidal forces indicates that the optimal configuration corresponds to a WH with mass $M \geq M_\odot\times10^{-2}$ and parameters $\lambda_0 \leq 10^{-2}$ and $\tau_0 \leq 10^{-3}$ approximately, thereby ensuring compliance with the WCCC. 
%
However, the most interesting consequence of the present work is that if the dilatonic interaction exists, that is, if there are phenomena where the scalar field is the source of the electromagnetic field of a celestial body, then the existence of traversable celestial WHs is possible. In fact, any theory that predicts extra dimensions, such as superstring theory \cite{Green:1984sg}, Kaluza-Klein theories \cite{Cita:KaluzaKleinComp},\cite{Kaluza:1921tu}, \cite{Klein:1926tv} or Brans-Dicke theory \cite{Brans:1961sx}, contains a dilaton field with exactly the interaction proposed here in Lagrangian (\ref{LagrangianoTesisUnidades}). In other words, the main result of the present work is that theories such as superstring theory, Kaluza-Klein or Brans-Dicke theory predict the existence of realistic celestial WHs, or the existence of celestial WHs in nature.


\section{Acknowledgements}

LB thanks SECIHTI-M\'exico for the doctoral grant.
This work was also partially supported by SECIHTI M\'exico under grants SECIHTI CBF-2025-G-1720 and CBF-2025-G-176. The authors are gratefully for the computing time granted by LANCAD and CONACYT in the Supercomputer Hybrid Cluster "Xiuhcoatl" at GENERAL COORDINATION OF INFORMATION AND COMMUNICATIONS TECHNOLOGIES (CGSTIC) of CINVESTAV. URL: http://clusterhibrido.cinvestav.mx/ and to Hector Oliver Hernandez for his help with the code installations.

\begin{appendices}
\section{Differential forms and the Hodge 2-form on the surface S} \label{ApnediceCargasInv}
We define the 2-surface:
\begin{equation}\label{eq:Sx_def}
S_x:\quad t=\text{const},\quad x=\text{const},\quad (y,\varphi)\in[-1,1]\times[0,2\pi).
\end{equation}
For large values of $|x|$, this surface is a topological 2-sphere in the asymptotic region (for both prolate and oblate charts), so the limit $x\to\infty$ (or $x\to-\infty$ for a second asymptotic extremum, when present) corresponds to integration over a large sphere at infinity.
Let $H$ now be an antisymmetric 2-form with covariant components $H_{\mu \nu}=-H_{\nu \mu}$. We adopt the usual convention
\begin{equation}\label{eq:Hodge_def}
(*H)_{\mu\nu}=\frac12\sqrt{-g}\,\epsilon_{\mu\nu\alpha\beta}\,H^{\alpha\beta},
\qquad
H^{\alpha\beta}=g^{\alpha\gamma}g^{\beta\delta}H_{\gamma\delta},
\end{equation}
where 
\begin{equation*}
    \sqrt{-g}=L_{\pm}^3\frac{e^{2k}}{f}(x^2 \pm y^2).
\end{equation*}
Using the orientation convention $\epsilon_{txy\varphi}=+1$, we can verify (using antisymmetry and appropriate index permutations) that $\epsilon_{y\varphi tx}=+1$. Therefore, from \eqref{eq:Hodge_def} we obtain $(*H)_{y\varphi}=\sqrt{-g}\,H^{tx}$. We observe that $g^{xx}=f e^{-2k} (x^2 \pm1)/(x^2 \pm y^2)L_{\pm}^2$ and $
g^{yy}=f e^{-2k} (1 -y^2)/(x^2 \pm y^2)L_{\pm}^2$, and thus we arrive at
\begin{equation}\label{eq:HodgeHMaestra}
(*H)_{y\varphi}=
L_{\pm}\, (x^2 \pm1)\Big(g^{tt}\,H_{tx}+g^{t\varphi}\,H_{\varphi x}\Big).
\end{equation}
Finally, if $\Omega$ is a 2-form, then
\begin{equation}\label{eq:Integral2FormaSobreSx}
    \int_{S_x}\Omega=\int_0^{2\pi}\!\!d\varphi\int_{-1}^{1}\!\!d y\;(\Omega_{y\varphi})\Big|_{t,x=\text{const}}.
\end{equation}

Therefore, when integrating $*H$ over $S_x$, we only need the component $( *H )_{y\varphi}$.
\section{NEC in the diagonal tetrad} \label{NEC in the diagonal tetrad}
By substituting metric functions of $\lambda_{c}$ (\ref{SolucionLambdaCombinada}) the NEC in $\mathbb{O}$ is (The upper symbol refers to WH, and the lower symbol refers to BH.)
{\setlength{\abovedisplayskip}{-10pt}
 \setlength{\abovedisplayshortskip}{-10pt}
\begin{widetext}
\begin{multline}\label{rho - varrho Lambdac}
    \varrho-P=\frac{ e^{-2k_c}}{2L_\pm ^2 (x^2\pm y^2)^5}   \bigg\{ 4\lambda_0 \tau_0 x y( x^2\pm y^2 ) \Big( x^2 \mp 1 \pm 2y^2+4k_0(x^2\pm 1) \Big)+ 2\lambda_0^2 \Big( y^4(1-y^2) +x^4(1+(1+8k_0)y^2 )
    \\ \pm 2(4k_0+y^2) x^2 y^2 \Big) + \tau_0^2 \Big[ x^2 y^2 \left(4 k_0 \left(y^2-2\right)-7 y^2+6\right)+\left(4 k_0+1\right) ( x^6\pm x^4 \left(2 y^2-1\right) \mp y^4 ) \Big]\bigg\}.
\end{multline}
\end{widetext}
}
\section{Riemann tensor components in an astronaut reference frame} \label{Riemman en astronautFrame}
The Riemann components of the solution (\ref{SolucionLambdaCombinada}) in $\overline{\mathbb{O}}$ are
{\small
{\setlength{\abovedisplayskip}{-10pt}
 \setlength{\abovedisplayshortskip}{-10pt}
\begin{subequations}\label{Riemman en SR del astronauta - LambdaCombinada}
\begin{equation}\label{R2121-Lambdac}
    \tensor{R}{_{ 1010 }}=\frac{ (1-y^2) e^{-2k_c}}{4L^2 (x^2+y^2)^5} \Big( \lambda_0(y^2-x^2) +2xy \tau_0 \Big)^2,
\end{equation}
\begin{equation}\label{R3131-Lambdac}
    \tensor{R}{_{ 2020 }} =\frac{e^{-2k_c}}{4 L^2 (x^2+y^2)^5 } \Big( \tau_0(x^2-y^2) +2xy \lambda_0 \Big)^2 ,
\end{equation}
\begin{multline}\label{R4141-Lambdac}
    \tensor{R}{_{ 3030 }} =\frac{e^{-2k_c}}{4 L^2 (x^2+y^2)^4}  \Bigg\{4\lambda_0 \tau_0 xy\Big( x^2-y^2 \Big)
    \\+\lambda_0^2 \Big( y^2(1-y^2)+x^2(3y^2+1) \Big) 
    \\+\tau_0^2 \Big( x^2(1+x^2-3y^2)+y^2 \Big) \Bigg\},
\end{multline}
\begin{equation}\label{R3232-Lambdac}
    \tensor{R}{_{ 2121 }} = 4 k_0 \tensor{R}{_{ 3030 }},
\end{equation}
\begin{multline}\label{R4242-Lambdac}
    \tensor{R}{_{ 3131 }} =  \frac{e^{-2k_c}}{4L^2 (x^2+y^2)^5}  
    \Bigg\{4 \lambda _0 \tau _0 x y \left(x^2-y^2\right) \\
    \Bigl( 4 k_0 \left(x^2-y^2+2\right)+3 \left(y^2-1\right)\Bigl)
    \\  +\lambda_0^2 \Bigg( x^4 \left(y^2(20 k_0-3)-4 k_0+3\right)
   \\ -2 x^2 y^2 \left(4 k_0 \left(y^2-3\right)-3
   y^2+3\right)
   \\+\left(4 k_0-3\right) y^4 \left(y^2-1\right)\Bigg)
   \\ +4 \tau _0^2 \bigg( k_0 \left(x^6
   +x^4 \left(1-2 y^2\right)+x^2 y^2 \left(5 y^2-6\right)+y^4\right)
   \\ +3 x^2 y^2 \left(1-y^2\right) \bigg)
    \Bigg\}.
\end{multline}
\end{subequations}
}}
\section{Electromagnetic vectorial field in oblates spheroidal coordinates}\label{EM Vectorial Field in Weyl}
{\small
{\setlength{\abovedisplayskip}{-10pt}
 \setlength{\abovedisplayshortskip}{-10pt}
\begin{subequations}\label{CampoElectromagneticoxy}
\begin{align}\label{CampoMagnetico}
    B_x&=  -\frac{L \, e^{-\lambda_c}}{2(x^2+y^2)^3} \bigg( \lambda _0^2 \, x \left(1-y^2\right) \left(x^2-y^2\right) \notag\\
    & +\tau _0 \left(x^2+1\right) \left(x^4+2 \tau _0 x y^2-y^4\right) \notag \\
    & + \lambda _0 \, y \Big(2 x \left(x^2+1\right) \left(x^2+y^2\right) \notag \\
    &+\tau _0 \left(y^2-x^4-3 x^2 \left(1-y^2\right) \right) \Big) \, \, \bigg).\\
    B_y&=   \frac{L \,e^{-\lambda_c}}{2(x^2+y^2)^3} \bigg( - 2 \lambda _0^2 x^2 y \left(1-y^2\right) \notag \\
    & +\tau _0 y \Big(\tau _0 \left(x^2+1\right) \left(x^2-y^2\right) \notag \\
    &-2 x \left(1-y^2\right) \left(x^2+y^2\right)\Big) \notag \\
    & + \lambda _0 \Big(\left(y^2-1\right) \left(y^4-x^4\right) \notag \\
    &+\tau _0 x \left(x^2 \left(3 y^2-1\right)-y^2 \left(y^2-3\right)\right)\Big) \,
    \bigg),
\end{align}

\begin{equation}\label{CampoElectrico}
    \begin{bmatrix} E_x \\ E_y \end{bmatrix}=\frac{e^{-\lambda_c}}{2(x^2+y^2)}\, 
    \begin{bmatrix} 2 \lambda_0 \, xy +\tau_0 (x^2 -y^2) \\ 2\tau_0 xy - \lambda_0(x^2-y^2) \end{bmatrix} ,
\end{equation}
\end{subequations}
}}
where $E$ represents the electric field and $B$ the magnetic field, measured in volts per meter and tesla, respectively.
\section{Table of dimension values}

{\small
\begin{table*}[t]
\caption{ Description of the WH parameters, using $\lambda_0 = 10^{-1}$ and $\tau_0 = 10^{-2}$. Here we focus on the SU-E scenario, assuming $L_+ \in \mathbb{R}$. We use $f_0=1=\kappa_0$, $l_1=r_s/2$, $J_\infty=-\lambda_0L_{+}^2/2$, $|\mathcal{M}|=\sqrt{l_1^2+N_\infty^2}$, $a=J_\infty/|\mathcal{M}|$ and
$N_\infty=Q_L=H_L=\tau_0 L_{+}/2$. Setting $\tau_0 = 0$ gives $N_\infty = Q_L = H_L = 0$, while $L_{+}, a,$ and $J_\infty$ remain essentially unchanged. For a comparison, refer to Table \ref{ValoresDeParametrosReferenciasReales}.}
\label{ValoresDeParametrosWH}
\begin{ruledtabular}
\begin{tabular}{c cccccc}
Size & 
\makecell{$l_1$\\(m)} &
\makecell{$L_+$\\(m)} &
\makecell{$J_\infty$\\(m$^2$)} &
\makecell{$a$\\(m)} &
\makecell{$N_\infty$\\(m)} &
\makecell{$Q_L$=$H_L$ \\(m)} \\
\hline
\text{Earth} &
$4.435\times 10^{-3}$ &
$8.9257\times 10^{-2}$ &
$-3.9834\times 10^{-4}$ &
$-8.9366\times 10^{-2}$ &
$4.4628 \times 10^{-4}$ & 
$4.4628 \times 10^{-4}$  \\
\text{Sun} &
$1.4766\times 10^{3}$ &
$2.9717\times 10^{4}$ &
$-4.4156\times 10^{7}$ &
$-2.9754\times 10^{4}$ &
$1.4859 \times 10^{2}$ & 
$1.4859 \times 10^{2}$  \\
\text{Magnetar/Pulsar} &
$2.0673\times 10^{3}$ &
$4.1606\times 10^{4}$ &
$-8.6551\times 10^{7}$ &
$-4.1656\times 10^{4}$ &
$2.080 \times 10^{2}$ & 
$2.080 \times 10^{2}$  \\
\text{Sgr A$^\ast$} &
$6.1309\times 10^{9}$ &
$1.2339\times 10^{11}$ &
$-7.6123\times 10^{20}$ &
$-1.2354\times 10^{11}$ &
$6.1694\times 10^{8} $ & 
$6.1694\times 10^{8} $\\
\end{tabular}
\end{ruledtabular}
\end{table*}
}

{\small
\begin{table*}[t]
\caption{ Description of the BH parameters, using $\lambda_0 = 10^{-1}$ and $\tau_0 = 10^{-2}$. Here we focus on the S-E scenario, assuming $L_- \in \mathbb{R}$. We use $f_0=1=\kappa_0$, $l_1=r_s/2$, $J_\infty=-\lambda_0L_{-}^2/2$, $|\mathcal{M}|=\sqrt{l_1^2+N_\infty^2}$, 
$a=J_\infty/|\mathcal{M}|$ and
$N_\infty=Q_L=H_L=\tau_0 L_{+}/2$. Setting $\tau_0 = 0$ gives $N_\infty = Q_L = H_L = 0$, while $L_{-}, a,$ and $J_\infty$ remain essentially unchanged. For a comparison, refer to Table \ref{ValoresDeParametrosReferenciasReales}.}
\label{ValoresDeParametrosBH}
\begin{ruledtabular}
\begin{tabular}{c cccccc}
 Size & 
\makecell{$l_1$\\(m)} &
\makecell{$L_+$\\(m)} &
\makecell{$J_\infty$\\(m$^2$)} &
\makecell{$a$\\(m)} &
\makecell{$N_\infty$\\(m)} &
\makecell{$Q_L$=$H_L$ \\(m)} \\
\hline
\text{Earth} &
$4.435\times 10^{-3}$ &
$4.4294\times 10^{-3}$ &
$-9.8099\times 10^{-7}$ &
$-2.2119\times 10^{-4}$ &
$2.2147 \times 10^{-5}$ & 
$2.2147 \times 10^{-5}$  \\
\text{Sun} &
$1.4766\times 10^{3}$ &
$1.4747\times 10^{3}$ &
$-1.0874\times 10^{5}$ &
$-7.3644\times 10^{1}$ &
$7.3737 $ & 
$7.3737$  \\
\text{Magnetar/Pulsar} &
$2.0673\times 10^{3}$ &
$2.0647\times 10^{3}$ &
$-2.1315\times 10^{5}$ &
$-1.0310\times 10^{2}$ &
$1.0321 \times 10^{1}$ & 
$1.0321 \times 10^{1}$  \\
\text{Sgr A$^\ast$} &
$6.1309\times 10^{9}$ &
$6.1232\times 10^{9}$ &
$-1.8747\times 10^{18}$ &
$-3.0577\times 10^{8}$ &
$3.0616\times 10^{7} $ & 
$3.0616\times 10^{7} $\\
\end{tabular}
\end{ruledtabular}
\end{table*}
}


\begin{table*}[t]
\caption{ Observational real Astrophysical reference values in geometric units (meters). We use
$l_1=r_s/2$, $J_\infty=GJ/c^3$, $|\mathcal{M}|=\sqrt{l_1^2+N_\infty^2}$ and
$a=J_\infty/|\mathcal{M}|$. We set $H_L= 0$ and $N_\infty=0$,
so $L_-=\sqrt{l_1^2-a^2-Q_L^2}$ is real.}
\label{ValoresDeParametrosReferenciasReales}
\begin{ruledtabular}
\begin{tabular}{c cccccccc}
 & 
\makecell{$l_1$\\(m)} &
\makecell{$L_-$\\(m)} &
\makecell{$J_\infty$\\(m$^2$)} &
\makecell{$a$\\(m)} &
\makecell{$Q_L$\\(m)}  \\
\hline
\text{Sun} &
$1.4766\times 10^{3}$ &
$1.4411\times 10^{3}$ &
$4.7560\times 10^{5}$ &
$3.2209\times 10^{2}$ & $0$ \\
\text{Magnetar} &
$2.0673\times 10^{3}$ &
$2.0219\times 10^{3}$ &
$7.7369\times 10^{2}$ &
$3.7426\times 10^{-1}$ & $\sim10^{4}$ \\
\text{Pulsar} &
$2.0673\times 10^{3}$ &
$1.8250\times 10^{3}$ &
$1.7948\times 10^{6}$ &
$8.6822\times 10^{2}$ & $\sim10^{2}$ \\
\text{Sgr A$^\ast$} &
$6.1309\times 10^{9}$ &
$2.6724\times 10^{9}$ &
$3.3830\times 10^{19}$ &
$5.5179\times 10^{9}$ & $0$ \\
\end{tabular}
\end{ruledtabular}
\end{table*}

\end{appendices}
\bibliography{Bibliografia}
\end{document}